\documentclass[12pt]{article}
 \usepackage{hyperref}
 \usepackage{graphicx}
  \usepackage{epsfig}
  \usepackage{amssymb}
  \usepackage{subfigure}
  \usepackage{amsmath} 

\evensidemargin - .5truecm
\allowdisplaybreaks[1]

\begin{document}

\newcommand{\lsim}{\lesssim}
\newcommand{\gsim}{\gtrsim}

\newcommand{\CC}{{\mathbb C}}
\newcommand{\RR}{{\mathbb R}}
\newcommand{\ZZ}{{\mathbb Z}}
\newcommand{\QQ}{{\mathbb Q}}
\newcommand{\NN}{{\mathbb N}}
\newcommand{\beq}{\begin{equation}}
\newcommand{\eeq}{\end{equation}}
\newcommand{\beal}{\begin{align}}
\newcommand{\eeal}{\end{align}}
\newcommand{\nn}{\nonumber}
\newcommand{\bea}{\begin{eqnarray}}
\newcommand{\eea}{\end{eqnarray}}
\newcommand{\ba}{\begin{array}}
\newcommand{\ea}{\end{array}}
\newcommand{\bfig}{\begin{figure}}
\newcommand{\efig}{\end{figure}}
\newcommand{\bc}{\begin{center}}
\newcommand{\ec}{\end{center}}

\newenvironment{appendletterA}
{
  \typeout{ Starting Appendix \thesection }
  \setcounter{section}{0}
  \setcounter{equation}{0}
  \renewcommand{\theequation}{A\arabic{equation}}
 }{
  \typeout{Appendix done}
 }
\newenvironment{appendletterB}
 {
  \typeout{ Starting Appendix \thesection }
  \setcounter{equation}{0}
  \renewcommand{\theequation}{B\arabic{equation}}
 }{
  \typeout{Appendix done}
 }

\begin{titlepage}
\nopagebreak

\renewcommand{\thefootnote}{\fnsymbol{footnote}}
\vskip 1cm

\vspace*{.5cm}
\begin{center}
{\Large \bf 
Heavy quark mass effects 
in the Energy-Energy Correlation in the back-to-back region
\\[0.1cm]
}
\end{center}


\par \vspace{1.5mm}
\begin{center}
  {\bf  Ugo Giuseppe\ Aglietti${}^{(a)}$}  and~ {\bf Giancarlo Ferrera${}^{(b)}$}\\

\vspace{5mm}

${}^{(a)}$
Dipartimento di Fisica, Universit\`a di Roma ``La Sapienza'' and\\ INFN, Sezione di Roma,
I-00185 Rome, Italy\\\vspace{1mm}

${}^{(b)}$ 
Dipartimento di Fisica, Universit\`a di Milano and\\ INFN, Sezione di Milano,
I-20133 Milan, Italy\\
\vspace{1mm}

\end{center}

\vspace{.5cm}


\par 
\begin{center} {\large \bf Abstract} \end{center}
\begin{quote}
\pretolerance 10000

We consider heavy quark mass ($m$) effects
in the Energy-Energy Correlation
function in $e^+e^-\mapsto\mathrm{hadrons}$
at high energy $Q$, 
in the back-to-back (two-jet) region.
In the ultra-relativistic limit, $Q \gg m$,
the QCD Sudakov
form factor $S(b)$ in impact parameter ($b$-)space 
reads:
\bea
\log S(b) &=& -  \int\limits_{m^2}^{Q^2} 
\frac{dk^2}{k^2}
\left\{
\log\left(\frac{Q^2}{k^2}\right) \,
A\left[\alpha_S\left(k^2\right)\right]
\, + \, 
B\left[\alpha_S\left(k^2\right)\right]
\right\}
\big[
1 \, - \, J_0\left(b k\right)
\big]
\nonumber\\
&&- \,\int\limits_0^{m^2} 
\frac{dk^2}{k^2}
\left\{
\log\left(\frac{Q^2}{m^2}\right) \,
A\left[\alpha_S\left(k^2\right)\right]
\, + \, 
D\left[\alpha_S\left(k^2\right)\right]
\right\}
\big[
1 \, - \, J_0\left(b k\right)
\big].
\nonumber
\eea
The double-log function 
$A(\alpha_S)$ describes
the effects of soft gluons,
quasi-collinear to the original quark-antiquark pair, 
while the single-log functions
$B(\alpha_S)$ and 
$D(\alpha_S)$
describe hard collinear radiation and soft
radiation
not log-collinearly enhanced, respectively.
The usual logarithmic expansion 
of the low transverse momenta contribution
to the form factor ($k^2<m^2$)
involves two new function series.
An explicit evaluation of the 
heavy quark form factor
at Next-to-Next-to-Leading Logarithmic accuracy is presented.
The relation of our results with the
well-known (classical) dead-cone effect in gauge theories is analyzed.
At Next-to-Leading Logarithmic (NLL) accuracy, an improved formula for the dead cone
is proposed; the soft term
produces an increase of the dead-cone opening angle of a factor $\sqrt{e} \simeq 1.64872$.
However, beyond NLL, a physical interpretation
of this kind breaks down.
The generalization of the above formula to the resummation
of shape variables
in $e^+e^-$ annihilation,
or to the resummation of the transverse momentum distributions 
in hadron collisions,
is also briefly discussed.

\vfill
\end{quote}
\vspace*{0.75cm}
\vspace*{\fill} 
\begin{flushleft}
December 2024
\vspace*{-1cm}
\end{flushleft}
\end{titlepage}

\renewcommand{\thefootnote}{\fnsymbol{footnote}}

\newpage


\section{Introduction}

In this note we consider heavy quark mass effects
\cite{Csikor:1983dt, Ali:1984gzn}
in the Energy-Energy Correlation (EEC)
function \cite{Basham:1978bw} in 
\beq
e^+ e^- \, \mapsto \, \mathrm{hadrons}
\eeq
at high energy.
A remarkable theoretical accuracy has presently been reached
in perturbative EEC calculations,
both fixed-order 
\cite{Richards:1982te, Richards:1983sr, Dixon:2018qgp,
DelDuca:2016csb, DelDuca:2016ily, Moult:2018jzp, Ebert:2020sfi, Duhr:2022yyp}
and resummed ones
\cite{Collins:1981uk, Collins:1981va,
Collins:1985xx, Kodaira:1981nh,
Kodaira:1982cr, Kodaira:1982az,
deFlorian:2004mp, Tulipant:2017ybb,
Kardos:2018kqj, Dixon:2019uzg, Aglietti:2024xwv}.
A variety of models to describe non-perturbative effects
in the EEC has also been proposed
\cite{Kodaira:1981np, Fiore:1992sa, Nason:1995np,
Dokshitzer:1995qm, Dokshitzer:1999sh, Schindler:2023cww, Kang:2023big, Lee:2024esz, Kang:2024dja}.
Mass effects in EEC has been also recently considered both in the back-to-back \cite{vonKuk:2024uxe,vonKuk:2023jfd} and in the opposite (forward) \cite{Craft:2022kdo} limit.
Furthermore, the EEC spectrum has been measured
with great accuracy at the $Z^0$ peak
both by the LEP experiments 
($\mathcal{O}(10^6)$ events)
and the SLD collaboration
\cite{DELPHI:1992qrr,L3:1992btq,OPAL:1991uui,OPAL:1993pnw,SLD:1994idb}. 
Therefore, given the present, 
both
theoretical and experimental status of the EEC, the consistent inclusion of (model-independent, perturbative) heavy quark mass effects
is particularly relevant.
Let us indeed consider, for concreteness sake, the following process:
\beq
\label{reaction_b-only}
e^+ e^- \, \to \, b \bar{b}.
\eeq
At a Center-Of-Mass (C.O.M.) energy $Q=30\, \mathrm{GeV}$, the simple estimate
\beq
\left( \frac{2 m_b}{Q} \right)^2 \, \approx \, 0.1 
\, \approx \, \alpha_S\left(Q^2\right), 
\eeq
suggests that beauty quark mass effects can substantially affect 
a leading-order (LO), $\mathcal{O}(\alpha_S)$, analysis of experimental data,
where $\alpha_S(Q^2)\simeq 0.14$  is the QCD coupling at $Q=30\, \mathrm{GeV}$.
By increasing the hard scale $Q$ to the $Z^0$ peak,
i.e. by taking $Q=m_Z$, beauty mass effects are expected to decrease
by an order of magnitude with respect to the previous
case, as:
\beq
\left( \frac{2m_b}{m_Z} \right)^2 \, \approx \, 0.01 
\, \approx \, \alpha_S^2\left(m_Z^2\right);
\eeq
where we have taken into account that
$\alpha_S(m_Z^2)\simeq 0.12$.
The above estimate suggests mass-related modifications of the EEC spectrum
at the NLO ($\mathcal{O}(\alpha_S^2)$) level.
Furthermore, at a future $e^+e^-$ accelerator 
with an C.O.M. energy of, let's say, $1\,\mathrm{TeV}$,
it would be interesting to study top-quark mass effects in the EEC 
function in
\beq
e^+ e^- \, \to t \, \bar{t}.
\eeq
Heavy quark mass effects have been 
studied in the framework of threshold resummation in general in \cite{Aglietti:2006wh} and in the context of heavy flavor decays in \cite{Aglietti:2007bp,Aglietti:2008xn,
Aglietti:2022rcm}.
Jets at LHC initiated by massive (beauty) quarks
have been extensively studied in \cite{Caletti:2023spr,Ghira:2023bxr,Gaggero:2022hmv,Caletti:2022hnc,Fedkevych:2022mid}.

The differential spectrum of the EEC function
is defined as \cite{Basham:1978bw,
Ellis:1996mzs, Marzani:2019hun}: 
\beq
\frac{d\Sigma^{\mathrm{EEC}}}{d \cos\chi}
\, \equiv \,
\sum_{N=2}^\infty
\sum_{i,j=1}^N
\frac{E_i E_j}{Q^2} \, d\sigma \, \delta(\cos\chi + \cos\theta_{ij})
\qquad (0 \le \chi \le \pi).
\eeq
A sum over any number $N=2,3,4,\cdots$ of hadrons (partons in perturbative QCD (pQCD)) in the final state is implied. 
The internal sum is over all pairs $(i,j)$ of hadrons
in the event and $\theta_{ij}$ is the (unoriented) angle between the 
space momenta ${\bf p}_i$ and ${\bf p}_j$ of the pair%
\footnote{
With our definition of the correlation angle $\chi$, the point
$\chi=0$ corresponds to the back-to-back
configuration, while the point $\chi=\pi$
corresponds to the forward region
(collinear pairs).
}.
Finally, $d\sigma$ is the differential two- and one-particle
inclusive cross section for $i\ne j$ and $i=j$
respectively.
In simple two-jet events, 
the EEC function describes the distribution
of the angular distance between them.
In a general complex, multi-jet event,
the EEC function describes the sum
of the distributions of the angular distance 
of all jet pairs.

The paper is organized as follows.
In section \ref{sec_main_res} we derive the main result,
which is actually quite simple and 
physically very intuitive, so let's sketch it.
Consider the Sudakov form factor in impact-parameter ($b$-)space \cite{Parisi:1979se,Curci:1979sk,Curci:1979bg},
in the usual massless case:
\beq
S_0(b) \, = \, \exp \, - \,\int\limits_0^{Q^2} 
\frac{dk_\perp^2}{k_\perp^2}
\left\{
\log\left(\frac{Q^2}{k_\perp^2}\right) \,
A\left[\alpha_S\left(k_\perp^2\right)\right]
\, + \, 
B\left[\alpha_S\left(k_\perp^2\right)\right]
\right\}
\Big[
1 \, - \, J_0\left(b k_\perp\right)
\Big]. 
\eeq
The double-log function 
$A(\alpha_S)=A_1\alpha_S+A_2 \alpha_S^2+\cdots$ describes
the effects of soft gluons
quasi-collinear to the original quark-antiquark pair, while the single-log function
$B(\alpha_S)=B_1\alpha_S+B_2\alpha_S^2+\cdots$ describes mostly hard collinear radiation (all the symbols are defined in detail in section \ref{sec_main_res}).
By going over the massive case, we find that
$S_0(b)$ is simply modified by replacing the 
lower integration limit on $k_\perp^2$, by $m^2$:
\beq
\label{eq_Shat_0}
S_0(b) \,\, \mapsto \,\, \hat{S}(b) 
\, = \, \exp \, - \int\limits_{m^2}^{Q^2} 
\frac{dk_\perp^2}{k_\perp^2}
\left\{
\log\left(\frac{Q^2}{k_\perp^2}\right) \,
A\left[\alpha_S\left(k_\perp^2\right)\right]
\, + \, 
B\left[\alpha_S\left(k_\perp^2\right)\right]
\right\}
\Big[
1 \, - \, J_0\left(b k_\perp\right)
\Big]. 
\eeq
The contribution from the cut or "missing" $k_\perp^2<m^2$ region
is "compensated" by adding to $\hat{S}(b)$ the following term:
\beq
\label{eq_DeltaS_b}
\Delta S(b) 
\, = \, \exp \, - \int\limits_0^{m^2} 
\frac{dk_\perp^2}{k_\perp^2}
\left\{
\log\left(\frac{Q^2}{m^2}\right) \,
A\left[\alpha_S\left(k_\perp^2\right)\right]
\, + \, 
D\left[\alpha_S\left(k_\perp^2\right)\right]
\right\}
\Big[
1 \, - \, J_0\left(b k_\perp\right)
\Big].  
\eeq
The single-log function
$D(\alpha_S)=D_1\alpha_S+D_2\alpha_S^2+\cdots$ describes soft radiation
not log-collinearly enhanced.
Note the similarities of the two above contributions to $S(b)$
and the fact that the leading 
function, namely $A(\alpha_S)$, occurs 
both in the high-$k_\perp$ term ($k_\perp^2>m^2$) and in the low-$k_\perp$ one
($k_\perp^2<m^2$).
Furthermore, by going from $\hat{S}(b)$ to $\Delta S(b)$, 
the infrared logarithm
\beq
\log\left(\frac{Q^2}{k_\perp^2}\right),
\eeq
of collinear origin,
is "frozen" into the large,
$k_\perp$-independent mass logarithm 
\beq
\log\left(\frac{Q^2}{m^2}\right) \, \gg \, 1.
\eeq
The above logarithm is large because, as already noted, we work in the ultra-relativistic limit
$Q^2 \gg m^2$.
Unlike $A(\alpha_S)$, the subleading functions,
$B(\alpha_S)$ and $D(\alpha_S)$,
are instead different (that is a quite common
situation in physics).
These functions describe complementary 
physical effects.
 
In section\,\ref{sec_funct_ser} we derive the usual function-series
representation for the QCD Sudakov form factor $S(b)$.
As well known, the order of the truncation of such series
defines the logarithmic accuracy: 
Leading-Logarithmic (LL) accuracy (keep only the first function of the series), 
Next-to-Leading-Logarithmic (NLL) accuracy (keep also the second function), 
Next-to-Next-to-Leading-Logarithmic  (N$^2$LL) accuracy  (the first three functions),
Next-to-Next-to-Next-to-Leading-Logarithmic  (N$^3$LL) accuracy  (the first four functions), and so on.
Also in the massive case, the results obtained have a very clear
physical picture.
The form factor $S(b)$ has a short-distance
(or high-energy) component $S_S(b)$,
defined for 
\beq
\label{eq_b_small}
b \, \lsim \, \frac{1}{m},
\eeq
basically coming from the integration over $k_\perp$
on the r.h.s. of eq.(\ref{eq_Shat_0}),
as well as a long-distance
(or low-energy) component $S_L(b)$,
defined for
\beq
b \, \gsim \, \frac{1}{m},
\eeq 
coming from the integration on the r.h.s. of eq.(\ref{eq_DeltaS_b})
(the precise definitions and formulae will be given
in section \ref{sec_funct_ser}).
It turns out $S_S(b)$ that is equal to the massless form factor,
the only difference being, 
in the massive case,
the restriction (\ref{eq_b_small}) 
 (which is clearly absent in the
massless case, for which $1/m=\infty$, so as to render
the inequality (\ref{eq_b_small}) trivial). 
The exponent of $S_S(b)$ involves a well-known function
series, which has recently been computed 
in pQCD
up to
N$^3$LL \cite{Aglietti:2024xwv}.
The long-distance form factor, in which mass effects
strongly manifest themselves, has instead an exponent
determined by $two$ function series,
coming from the integration of the
$A$-type and $D$-type terms on the r.h.s. of 
eq.(\ref{eq_DeltaS_b}).
The first series is multiplied by
$\log(Q^2/m^2)$ and describes the leading
(double-logarithmic) infrared effects while
the second series describes subleading effects
related to soft radiation not log-collinearly
enhanced.
A logarithmic accuracy scheme for
these two series, namely for their consistent truncations, 
is constructed. 
We also find that it is also possible to combine them
into a single series. 
 
In section\,\ref{sec_res_NLL} we apply the general scheme
built up in the previous section,
to explicitly evaluate the massive form factor
$S(b)$ at Next-to-Next-to-Leading Logarithmic 
accuracy for all the series involved,
N$^2$LL\,+\,N$^2$LL$_m$
(we keep the first three functions
in each series). 
An intriguing result is that,
at the Next-to-Leading level,
the massive form factor 
can be simply evaluated, by imposing to the
massless form factor an "improved"
dead-cone\cite{Dokshitzer:1991fd,Dokshitzer:1995ev} restriction, given by:
\beq
\theta \, \gsim \, \frac{\sqrt{e}}{\gamma},
\eeq
where 
\beq
\gamma \, \equiv \,\frac{E}{m} \, = \, 
\frac{1}{\sqrt{1 \, - \, v^2}} \, \gg \, 1
\eeq
is the usual Lorenz factor of the (ultra-relativistic) massive quark
(with $E$ the quark energy and $v$ its ordinary 3-velocity).
The improvement is provided by
introduction of the factor $\sqrt{e} \simeq 1.64872$,
with $e\simeq 2.71828$ the Napier constant \cite{Dokshitzer:2005ri}.
In general, we find that the dead-cone
effect , originally discovered in classical
electrodynamics \cite{Jackson:1998nia}, 
receives quantum
corrections in QCD at each order of
perturbation theory.
However, a simple physical picture as the 
one sketched above,
does not extend beyond Next-to-Leading level.
At Next-to-Next-to-Leading level, for example,
two different quantum
corrections to the dead-cone opening angle
are found.

In section\,\ref{sec_phenomen} we discuss the numerical impact
of our N$^2$LL results for a hard scale $Q=30\,\mathrm{GeV}$, 
or at the $Z^0$ peak, 
i.e. for a hard scale $Q=m_Z$, 
in the case of beauty quarks. 
Plots both in the impact-parameter ($b$-)space
and in the physical (angle-$\chi$) space
are provided.
%

Finally, section\,\ref{sec_concl} contains the conclusion
of our analysis, together with a discussion
of natural generalizations of the results.
Our results indeed can be improved along
many different directions.


\section{Resummation in the back-to-back region for $m \ne 0$}
\label{sec_main_res}

In the back-to-back (or two jet) region, the EEC distribution
is conveniently written in the following factorized form\,\cite{Catani:1992ua}:
\beq
\frac{1}{ \sigma_t\left(\alpha_S\right) } \,
\frac{d\Sigma^{\mathrm{EEC}}}{d y}\left(y;\alpha_S\right)
\, = \, \frac{1}{2} H(\alpha_S) \, {\cal S} (y;\alpha_S) 
\, + \, \mathrm{Rem}(y;\alpha_S);
\eeq
where $\alpha_S \equiv \alpha_S\left(Q^2\right)$ is the QCD coupling
at the C.O.M. energy $Q$ and we introduced the variable
\beq
y \, \equiv \, \frac{1 \, - \, \cos\chi}{2}
\qquad (0 \le y \le 1).
\eeq
Let us define the quantities entering
the above formula:
\begin{enumerate}
\item
${\cal S}(y;\alpha_S)$ is a universal, i.e. process-independent,
long-distance dominated QCD form factor, factorizing large double  logarithms of infrared nature (Sudakov logarithms),
occurring at each order of perturbation theory;
\item
$H(\alpha_S)$ is a process-dependent,
short-distance dominated Hard factor,
or Coefficient function,
which can be computed in fixed-order
(truncated) perturbation theory;
\item
$\mathrm{Rem}(y;\alpha_S)$
is a process-dependent,
short-distance dominated function of $y$,
which also can be computed in fixed-order
perturbation theory.
Its inclusion in the factorization formula above
allows for a good description of the EEC spectrum
also outside the two-jet region.
\end{enumerate}
We normalize the EEC distribution
to the total, i.e. fully-inclusive, radiatively-corrected 
cross section $\sigma_t(\alpha_S)$.
In this paper we concentrate on the Sudakov
form factor, which contains the dominant contributions
in the back-to-back region and is thus the most sensitive quantity to
heavy quark mass effects,
as the latter enters 
the argument of the large infrared logarithms;
The evaluation of mass effects in the
Coefficient function and in the
Remainder function is left to future work.

The Sudakov form factor in physical (angle) space, ${\cal S}={\cal S}(y)$,
is obtained by means of a (numerical)
inverse Fourier-Bessel transform
of the Form factor $S = S(b)$ in impact-parameter ($b$-)space:
\beq
{\cal S}(y) \, = \, \int\limits_0^\infty d(Qb) \, \frac{Qb}{2}
\, J_0\left( b Q \sqrt{y} \right) \, S(b) .
\eeq
$J_0(x)$ is the Bessel function of first kind
(smooth at $x=0$), with zero index.
To simplify notation, the dependence on $\alpha_S$
in the form factors above has been omitted.
In  order to avoid the Landau singularity
occurring in $S(b)$ at a large impact-parameter value, $b=b_{LP}$, the
above integral can be cut at $b_{\mathrm{MAX}} < b_{LP}$,
for example at $b_{\mathrm{MAX}} = b_{LP}/2$
(see later).
More refined prescriptions can also be implemented
\cite{Catani:1996yz}.

In turn,
the Sudakov form factor in $b$-space is
given by the exponential of minus the (effective) one soft-gluon rate,
\beq
S(b) \, = \, \exp\big[ - \, w(b) \big].
\eeq
The $b$-space rate, $w=w(b;Q;m)$,
is given in the quasi-collinear limit 
\cite{Keller:1998tf,Catani:2000ef,Catani:2002hc},
to first order in $\alpha_S$, by: 
\beq
w \,= \, \frac{C_F}{\pi} \int\limits_0^{Q^2} 
dk_\perp^2 \, \alpha_S(k_\perp^2)
\int\limits_{k_\perp/Q}^1  
\frac{dz}{k_\perp^2 + z^2 m^2}
 \left[
\frac{2}{z} - 2 + z
- \frac{2 m^2 z (1-z) }{k_\perp^2 \, + \, z^2 m^2 }
\right]
\left[
1 \, - \, J_0\left(b k_\perp\right) 
\right].
\eeq
The quasi-collinear limit is defined
by: $k_\perp\to 0^+$, $m\to 0^+$ with
$k_\perp/m\to \mathrm{constant}$
(the hard scale $Q$ and $z$ are held constant)%
\footnote{
The quasi-collinear limit is a generalization to massive quarks
of the well-known
collinear limit for massless partons,
$k_\perp\mapsto 0^+$.
}.
As usual, $C_F=(N_C^2-1)/(2N_C) = 4/3$ for $N_C=3$ colors
in QCD, 
$z$ is the gluon energy fraction
(in the C.O.M. frame);
\beq
z \, \equiv \, \frac{2 E_g}{Q} 
\qquad (0 \le z \lsim 1);
\eeq
with $Q^2 \equiv (p_{e^-}^\mu+p_{e^+}^\mu)^2$ the hard scale squared.
Finally, $\bf{k}_\perp$ is the soft-gluon
transverse momentum, lying in the $xy$-plane, 
with length
\beq
k_\perp \, \equiv \, 2 E_g \sin\left(\frac{\theta}{2}\right) \, \simeq \, E_g \, \theta
\qquad\mathrm{for} \quad \theta \ll 1.
\eeq
Finally, $\theta$ is the angle between the space
momentum $\vec{k}$ of the final gluon and the $z$-axis of the space frame
$(0 \le \theta \le \pi)$.
By integrating over $z$ and neglecting
$\mathcal{O}\left(k_\perp\right)$ 
non-logarithmic, short-distance contributions
(to be included later in the Coefficient function
$H(\alpha_S)$
or in the Remainder function
$\mathrm{Rem}(y,\alpha_S)$, through matching
with exact, fixed-order calculations of the EEC spectrum), one obtains:
\bea
w \, = &+&
A_1 \int\limits_{m^2}^{Q^2} 
\frac{dk_\perp^2}{k_\perp^2} \, \alpha_S(k_\perp^2)
\log\left(\frac{Q^2}{k_\perp^2}\right) 
\left[
1 \, - \, J_0\left(b k_\perp\right)
\right] +
\nonumber\\
&+& B_1 \int\limits_{m^2}^{Q^2} 
\frac{dk_\perp^2}{k_\perp^2} \, \alpha_S(k_\perp^2)
\left[
1 \, - \, J_0\left(b k_\perp\right)
\right] +
\nonumber\\
&+& \left[
A_1\log\left(\frac{Q^2}{m^2}\right) 
+ D_1 \right]
\int\limits_0^{m^2}
\frac{dk_\perp^2}{k_\perp^2} \, \alpha_S(k_\perp^2)
\left[
1 \, - \, J_0\left(b k_\perp\right)
\right] ;
\eea
where we have introduced the standard first-order
coefficients
\bea
\label{abd1}
A_1 &=& + \frac{C_F}{\pi};
\nonumber\\ 
B_1 &=& - \frac{3}{2}\frac{C_F}{\pi};
\nonumber\\
D_1 &=& - \frac{C_F}{\pi}. 
\eea
The accuracy of resummation
is improved, as usual, by adding higher-order terms
to the first-order terms above,
so that the formally-exact rate is written:
\bea
\label{eq_main_resumm}
w \, = &+&
\int\limits_{m^2}^{Q^2} 
\frac{dk_\perp^2}{k_\perp^2} \, 
\log\left(\frac{Q^2}{k_\perp^2}\right)
A\left[\alpha_S(k_\perp^2)\right]
\left[
1 \, - \, J_0\left(b k_\perp\right)
\right] \, +
\nonumber\\
&+& \int\limits_{m^2}^{Q^2} 
\frac{dk_\perp^2}{k_\perp^2} \, 
B\left[ \alpha_S\left(k_\perp^2\right) \right]
\left[
1 \, - \, J_0\left(b k_\perp\right)
\right] \, +
\nonumber\\
&+&
\log\left(\frac{Q^2}{m^2}\right) 
\int\limits_0^{m^2}
\frac{dk_\perp^2}{k_\perp^2} \, 
A\left[\alpha_S(k_\perp^2)\right]
\left[
1 \, - \, J_0\left(b k_\perp\right)
\right] \, + 
\nonumber\\
&+& 
\int\limits_0^{m^2}
\frac{dk_\perp^2}{k_\perp^2} \, 
D\left[ \alpha_S(k_\perp^2) \right]
\left[
1 \, - \, J_0\left(b k_\perp\right)
\right];
\eea
where:
\bea
A\left(\alpha_S\right) &=& \sum_{n=1}^\infty
A_n \, \alpha_S^n;
\nonumber\\
B\left(\alpha_S\right) &=& \sum_{n=1}^\infty
B_n \, \alpha_S^n;
\nonumber\\
D\left(\alpha_S\right) &=& \sum_{n=1}^\infty
D_n \, \alpha_S^n.
\eea
The coefficients $A_i$ 
\cite{Kodaira:1982cr, Moch:2004pa, Becher:2010tm,
Moch:2018wjh, vonManteuffel:2020vjv, Li:2016ctv},
$B_i$ \cite{Kodaira:1982az,deFlorian:2004mp,Aglietti:2024xwv}
and $D_i$ \cite{Korchemsky:1987wg}, 
like the first-order ones, are constants,
i.e. they depend only on $N_C$ and $n_f$
(a logarithmic dependence of the coefficients
would imply indeed an incomplete factorization).
Let us stress that eq.(\ref{eq_main_resumm}) is one of the main results
of the paper.
The following remarks are in order:
\begin{enumerate}
\item
In the massless limit, $m \to 0^+$,
the lower limit of the integration
over $k_\perp^2$ in the first two terms 
on the r.h.s. of eq.(\ref{eq_main_resumm})
approaches zero, while
the third and fourth
terms vanish, so one recovers the well-known
massless rate:
\bea
\lim_{m \to 0^+} w
&=&
\int\limits_0^{Q^2} 
\frac{dk_\perp^2}{k_\perp^2} \, 
\log\left(\frac{Q^2}{k_\perp^2}\right)
A\left[\alpha_S\left(k_\perp^2\right)\right]
\left[
1 \, - \, J_0\left(b k_\perp\right)
\right] \, +
\nonumber\\
&+& \int\limits_0^{Q^2} 
\frac{dk_\perp^2}{k_\perp^2} \, 
B\left[ \alpha_S\left(k_\perp^2\right) \right]
\left[
1 \, - \, J_0\left(b k_\perp\right)
\right]. 
\eea
Therefore eq.(\ref{eq_main_resumm}) provides a {\it sensible}
generalization of the massless case.
Let us note that the integrands of the first two terms
on the r.h.s. of eq.(\ref{eq_main_resumm})
are {\it exactly} equal to the corresponding ones
of the massless case ($m=0$);
in other words, the {\it only} difference relies in the lower integration
limit on $k_\perp^2$, which is $m^2$ in the present case, 
while it is zero in the massless case.
\item
Even though we have derived eq.(\ref{eq_main_resumm}) 
for the EEC function in the back-to-back region,
this equation actually describes
heavy quark mass effects in general $q_T$-resummation, as well as in
related observables.
In other words, eq.(\ref{eq_main_resumm}) provides a general 
resummation scheme for massive quarks, which can be applied
to different cases.
For concreteness sake,
think, for example, to the heavy quark mass effects
in the wide or total jet broadening 
in $e^+e^-$ annihilation to hadrons
at high energy.
To implement massive resummation in the process under consideration,
one has only to insert on the r.h.s. of eq.(\ref{eq_main_resumm})
the $A_i$, $B_i$ and $D_i$ coefficients 
specific to that process.
Such claimed generality of eq.(\ref{eq_main_resumm})
clearly stems from its derivation,
as we have simply evaluated
the effective one-gluon rate in impact-parameter space $(k_\perp \mapsto b)$. 
We expect the structure of mass corrections
to be the same also in processes involving
massive quarks in the initial state,
such as $q_T$ lepton-pair, $W$ or $Z$ spectra in Drell-Yan processes \cite{Berge:2005rv,Pietrulewicz:2017gxc}.
\item
In the third term on the r.h.s. of eq.(\ref{eq_main_resumm}),
the hard scale $Q$ only enters
in the constant $\log(Q^2/m^2)$ term, which is outside
the integral over $k_\perp^2$,
while in the last term, the forth one, $Q$ does
not appear at all.
Therefore, the low-$k_\perp^2$ component ($k_\perp^2<m^2$)
of the rate is at most {\it linear} in $\log(Q^2/m^2)$,
i.e. no terms proportional, for example, to 
$\log^2(Q^2/m^2)$
or $\log^3(Q^2/m^2)$ do occur in $w(b)$.
A similar result has been obtained
in the Wilson-line
analysis of the soft behavior of the massive
quark form factor \cite{Korchemsky:1987wg}.
The cusp anomalous dimension, explicitly computed
up to second order in $\alpha_S$, contains,
in the ultra-relativistic limit, a linear term in
$\log(Q^2/m^2)$, together with a constant term.
In the same paper, an all-order proof 
in $\alpha_S$
of this linearity property is also provided.
\item
All the terms on the r.h.s. of eq.(\ref{eq_main_resumm}),
with the exception of the last term, involving
the function $D(\alpha_S)$, can be obtained simply by
imposing the dead-cone restriction $\theta \gsim m/E$
to the standard massless resummation formula
($\theta$ is the gluon emission angle and $E\gg m$ 
is the heavy-quark energy).
The $D$-term cannot be obtained by means of the dead cone,
for the good reason that it simply does not occur
in the massless distribution.
\item
The long-distance contribution to the one-gluon rate, 
$w=w(b)$ --- the last two terms on the r.h.s. of 
eq.(\ref{eq_main_resumm}) ---, can also be written in the 
following equivalent form, involving a single
series (rather than the two usual series 
involving the $A_n$'s
and $D_n$'s above):
\bea
&& \int\limits_0^{m^2} 
\frac{dk_\perp^2}{k_\perp^2}
\left\{
\sum_{n=1}^\infty
A_n \log\left[ 
\frac{Q^2 \exp\left( D_n/A_n \right) }{m^2} \right] 
\, \alpha_S^n\left(k_\perp^2\right)
\right\}
\Big[
1 \, - \, J_0\left(b k_\perp\right)
\Big] \, =
\nonumber\\
&=&
+ \, A_1 \, \log\left[ 
\frac{Q^2 \exp\left( D_1/A_1 \right) }{m^2} \right] 
\, \int\limits_0^{m^2} \frac{dk_\perp^2}{k_\perp^2} \, \alpha_S\left(k_\perp^2\right) \, +
\nonumber\\
&& \,\,  + \, A_2 \, \log\left[ 
\frac{Q^2 \exp\left( D_2/A_2 \right) }{m^2} \right] 
\, \int\limits_0^{m^2} 
\frac{dk_\perp^2}{k_\perp^2} \, \alpha_S^2\left(k_\perp^2\right)
\, + \, \cdots .
\eea
One has then to insert, inside
the powers of $\alpha_S\left(k_\perp^2\right)$,
its explicit asymptotic expansion
up to the required order.
\item
For a heavy quark, i.e. a quark with a mass $m \gg \Lambda_{QCD}$,
where $\Lambda_{QCD}$ is the QCD scale,
the short-distance contributions $(k_\perp^2>m^2)$
to the one-gluon rate
are well defined,
while the long distance ones $(k_\perp^2<m^2)$ are not, because
of the integration over the Landau singularity
of the QCD running coupling (a simple pole in lowest order).
A (non-perturbative) prescription is then needed to actually
compute the long-distance component
of the form factor.
\end{enumerate}


\section{Function-series representation}
\label{sec_funct_ser}

By integrating over $k_\perp^2$ 
the terms on the r.h.s. of eq.(\ref{eq_main_resumm}),
the Sudakov form factor in $b$-space, $S = S(b)$, can be written 
in terms of the the usual 
function-series expansions as follows:
\beq
\label{eq_def_Phi}
S(b) \, \equiv \, 
\left\{
\begin{array}{rl}
S_S(b) & \mbox{if} \,\,\,\,\, b \, \le \, b_{\mathrm{CR}} ;
\\
S_L(b) & \mbox{if} \,\,\,\,\, b \, > \, b_{\mathrm{CR}} ;
\end{array} 
\right. 
\eeq
The parameter
\beq
b_{\mathrm{CR}} \, \equiv \, \frac{b_0}{m}
\eeq
plays the role of a {\it critical length}, 
below which mass effects are neglected and
the short-distance ($S$) form factor $S_S$ occurs:
\beq
\label{eq_def_Phi_S}
S_S \, \equiv \, \exp\Big\{
L \, g_1(\lambda) \, + \, G\left[\lambda; \alpha_S\left(Q^2\right)\right]
\Big\};
\eeq
where:
\beq
\label{eq_def_F}
G\left(\lambda,\alpha_S\right)
\, \equiv \,
\sum_{n=0}^\infty
\alpha_S^n \, g_{n+2}(\lambda)
\, = \, g_2(\lambda) \, + \, \alpha_S \, g_3(\lambda)
\, + \, \alpha_S^2 \, g_4(\lambda) \, + \, \cdots . 
\eeq
As usual, we have defined:
\beq
b_0 \, \equiv \, 2 \exp\left(-\gamma_E\right) \, \simeq \, 1.123;
\eeq
with $\gamma_E\simeq 0.577$ the Euler constant%
\footnote{
$b_0$ is a dimensionless constant of order one,
ubiquitous in $q_T$-resummation and related resummations.
}.
At larger distances, $b > b_{\mathrm{CR}}$,
mass effects are instead substantial and
a different, larger distances dominated form factor $S_L$ 
comes into play:
\beq
\label{eq_def_S_L}
S_L \, \equiv \, \overline{S}_S \, 
\exp\left\{
\log\left( \frac{Q^2}{m^2} \right) 
\, F\left[ \rho; \alpha_S\left(m^2\right) \right]
\, + \, H\left[ \rho; \alpha_S\left(m^2\right) \right]
\right\};
\eeq
where:
\bea
\label{eq_def_G_and_H}
F\left( \rho; \alpha_S \right)
&\equiv&
\sum_{n=0}^\infty
\alpha_S^n \, f_{n+1}(\rho)
\, = \, f_1(\rho)\, + \, \alpha_S \, f_2(\rho)
\, + \, \alpha_S^2 \, f_3(\rho) \, + \, \cdots ;
\nonumber\\
H\left( \rho; \alpha_S \right)
&\equiv&
\sum_{n=0}^\infty
\alpha_S^n \, h_{n+2}(\rho)
\, = \, h_2(\rho)\, + \, \alpha_S \, h_3(\rho)
\, + \, \alpha_S^2 \, h_4(\rho) \, + \, \cdots.
\eea 
We have defined the variables:
\bea
\label{eq_def_lambda_and_rho}
\lambda &\equiv& 
\beta_0 \, \alpha_S\left(Q^2\right) \, L ;
\nonumber\\
\rho &\equiv& 
\beta_0 \,\alpha_S\left(m^2\right) \, \mathcal{L};
\eea
where:
\bea
L &\equiv& 
\log\left( \frac{Q^2 \, b^2}{b_0^2} \right);
\nonumber\\
\mathcal{L} &\equiv& \log\left(\frac{m^2 \, b^2}{b_0^2} \right).
\eea
Finally, we have defined the constant
\beq
\overline{S}_S \, \equiv \, S_S\left( b \mapsto b_{\mathrm{CR}} \right) 
\, = \, S_S\left[ L \mapsto \log\left( \frac{Q^2}{m^2} \right) ; \, \lambda \mapsto \beta_0 \, \alpha_S\left(Q^2\right) 
\, \log\left( \frac{Q^2}{m^2} \right) \right].
\eeq
The following remarks are in order:
\begin{enumerate}
\item
The short-distance form factor $S_S$,
defined in eq.(\ref{eq_def_Phi_S}),
is equal to the standard $b$-space Sudakov form factor
for the EEC in the massless case.
That is equivalent to say that, as far as resummation is concerned,
mass effects only manifest themselves {\it above}
a critical (transverse) distance $b_{\mathrm{CR}}$;
below such a distance, mass effects are {\it totally} absent.
The constant $b_{\mathrm{CR}}$
is basically the Compton wavelength 
\beq
\lambda_C \, \equiv \, \frac{\hbar}{m c} 
\eeq
of the heavy quark
(even though we usually take units for which 
$\hbar=c=1$, in this case, for clarity's sake, we have written $\lambda_C$ for general units). 
In quantum field theory, $\lambda_C$
has the general physical meaning
of minimal length for the localization of a particle of mass $m$; indeed,
if one tries to localize the particle
at smaller distances, additional
particle-antiparticle pairs are created
in the measuring process,
making the identification of the original
particle impossible \cite{Landau4}.
Therefore it should not come as a surprise 
the fact that $b_{\mathrm{CR}}$ sets the 
point of the {\it transition},
in $b$-space, from the short-distance region
$(b \, \le \, b_{\mathrm{CR}})$,
to the long-distance one $(b \, > \, b_{\mathrm{CR}})$.
\item
The long-distance form factor $S_L$ contains the
$b$-independent, large collinear logarithm
\beq
\label{eq_define_LCL}
\log\left( \frac{Q^2}{m^2} \right) \, = \, L \, - \, \mathcal{L}. 
\eeq
The above logarithm is large
because, as already remarked,
in the quasi-collinear limit, 
$Q^2 \gg m^2$.
This logarithm is actually larger than $L$, as
\beq
\log\left( \frac{Q^2 \, b^2}{b_0^2} \right)
\, \le \, \log\left( \frac{Q^2}{m^2} \right)
\quad \mathrm{for} \quad b \, \le \, b_{\mathrm{CR}}.
\eeq
The occurrence of the mass logarithm defined eq.(\ref{eq_define_LCL}),
inside $S_L$,
is related to the fact that, in the low-$k_\perp^2$
region ($k_\perp^2<m^2$), gluon emission angles are generally
smaller than in the  large-$k_\perp^2$ region
(we have integrated over the gluon energy $z$).
That implies that,
in the low-$k_\perp^2$ region, 
the dead-cone restriction becomes
relevant and the collinear singularity
of the QCD matrix elements squared
is screened by the heavy quark mass,
so the mass logarithm in eq.(\ref{eq_define_LCL}) occurs.
On the contrary, in the large-$k_\perp^2$ region, 
the dead-cone restriction is
irrelevant and the collinear singularity
is basically regulated by $1/b$, giving rise then to a $b$-dependent collinear logarithm.
\item
The constant $\overline{S}_S$ ensures the {\it continuity} of the form factor
$S=S(b)$
across the point $b = b_{\mathrm{CR}}$,
where its definition changes from one analytic expression
to another one (see eq.(\ref{eq_def_Phi})).
That is because the functions $F(\rho,\alpha_S)$ and $H(\rho;\alpha_S)$
do vanish for vanishing $\rho$ and
\beq
\lim_{b \, \to \, b_{\mathrm{CR}}^-} \log\left( \frac{Q^2 \, b^2}{b_0} \right) \, = \, \log\left( \frac{Q^2}{m^2} \right) ;
\qquad
\lim_{b \, \to \, b_{\mathrm{CR}}^+} \log\left(\frac{m^2 \, b^2}{b_0^2} \right) \, = \, 0.
\eeq
Note, however, that the transition is not
{\it smooth}; it can be made smooth,
i.e. infinitely differentiable ($C^\infty$), by means of the so-called partitions
of unity \cite{bo}.
\item
The long-distance form factor $S_L$
is determined by two different functions:
\begin{enumerate}
\item
The leading function $F(\rho;\alpha_S)$, 
multiplied by the large logarithm $\log(Q^2/m^2)$,
describing double-logarithmic effects;
\item
The subleading or "residual" function $H(\rho;\alpha_S)$,
originating from soft gluons not log-collinearly enhanced.
The first term of its series expansion,
$h_1(\rho) \equiv 0$ because,
at leading, double-log level, $\log(Q^2/m^2)$ necessarily occurs
at the exponent of $S_L$.
\end{enumerate}
\item 
The long distance form factor $S_L$ takes into account the (logarithmic) mass corrections
  from the massive quark legs (the so-called {\itshape primary effects}). In order to take into account
  {\itshape secondary effects} from a massive (real or virtual) quark pair, we have to remove a massless
  quark pair from the region $k_\perp<m$. That way, the number of active flavors in the coefficients entering the functions $F(\rho;\alpha_S)$
  and $H(\rho;\alpha_S)$ are reduced by one unit.
\item
Even though the infrared logarithms $L=L(b)$ 
and $\mathcal{L}=\mathcal{L}(b)$
both depend on the same variable,
namely the impact parameter $b$, they can be considered
independent variables, because theirs domain
of definition are disjoint, i.e. not overlapping
(they are $b \le b_{\mathrm{CR}}$ and $b>b_{\mathrm{CR}}$
respectively). 
That implies, in particular, that
one can use different logarithmic 
approximation schemes
for $S_S$ and $S_L$.
For instance, one can use a N$^3$LL
approximation for $S_S$ and a N$^2$LL
approximation for $S_L$.
One has just to take care of computing the constant
$\overline{S}_S$ to the same approximation
of $S_S$, in order to preserve
the continuity of the form factor at the transition point $b=b_{\mathrm{CR}}$.
For clarity's sake, from now on,
we will indicate the LL accuracy
for the long-distance form factor $S_L$
(evaluate only $f_1$)
with LL$_m$, the NLL accuracy 
(evaluate also $f_2$ and $h_2$)
with
NLL$_m$, and so on.
According to these new conventions,
the logarithmic accuracy of the above example
is then fully specified as:
\beq
\textrm{N}^3\textrm{LL} + 
\textrm{N}^2\textrm{LL}_m.
\eeq
\item
The definition of the function-series expansion for $F(\rho;\alpha_S)$
(the first of eqs.(\ref{eq_def_G_and_H}))
is slightly different from the (usual) definition
of the series expansions
of $G(\lambda;\alpha_S)$  (eq.(\ref{eq_def_F}))
or of $H(\rho;\alpha_S)$ (the second of eqs.(\ref{eq_def_G_and_H})).
That is related to the fact
that the function $F(\rho;\alpha_S)$, so all its
terms, are multiplied by the large logarithm
$\log(Q^2/m^2)$.
Indeed the $F$-contribution in eq.(\ref{eq_def_S_L})
can also be written in the "standard" form:
\beq
\log\left( \frac{Q^2}{m^2} \right) \, f_1(\rho) 
\, + \,
\alpha_S\left(m^2\right) \log\left( \frac{Q^2}{m^2} \right)
\sum_{n=0}^\infty
\alpha_S^n \, f_{n+2}(\rho);
\eeq
in which we have introduced the $\mathcal{O}(1)$
quantity  
\beq
\alpha_S\left(m^2\right) \log\left( \frac{Q^2}{m^2} \right).
\eeq
\item
The dependence of the functions $f_n(\rho)$
on the double-log coefficients $A_k$
can be made explicit by writing:
\bea
\label{eq_fn_Ak}
f_n(\rho) &=& \sum_{k=1}^n A_k \, t_k^{(n)}(\rho); 
\qquad n=1,2,3,\cdots;
\eea
where the $t_k^{(n)}(\rho)$'s are functions
of $\rho$ independent on the $A_k$'s.
For the first three functions,
the above expansion explicitly reads:
\bea
f_1(\rho) &=&  A_1 \, t_1^{(1)}(\rho);
\nonumber\\
f_2(\rho) &=& A_2 \, t_2^{(2)}(\rho) \, + \,
A_1 \, t_1^{(2)}(\rho); 
\nonumber\\
f_3(\rho) &=&
A_3 \, t_3^{(3)}(\rho) \, + \,
A_2 \, t_2^{(3)}(\rho) \, + \,
A_1 \, t_1^{(3)}(\rho). 
\eea
In a similar way,, 
the dependence of the subleading functions
$h_n(\rho)$ on the soft coefficients $D_k$
can be extracted out by writing
($h_1(\rho) \equiv 0$):
\bea
\label{eq_hnp1_Dk}
h_{n+1}(\rho) &=& 
\sum_{k=1}^n D_k \, t_k^{(n)}(\rho) 
\qquad (n=1,2,3,\cdots).
\qquad
\eea
For concreteness sake, for the first three functions, we have:
\bea
h_2(\rho) &=&  D_1 \, t_1^{(1)}(\rho);
\nonumber\\
h_3(\rho) &=& D_2 \, t_2^{(2)}(\rho) \, + \,
D_1 \, t_1^{(2)}(\rho); 
\nonumber\\
h_4(\rho) &=&
D_3 \, t_3^{(3)}(\rho) \, + \,
D_2 \, t_2^{(3)}(\rho) \, + \,
D_1 \, t_1^{(3)}(\rho). 
\eea
The crucial point is that the functions $t_k^{(n)}(\rho)$
are the same in both equations (\ref{eq_fn_Ak}) and (\ref{eq_hnp1_Dk}), for any $n$ and $k$.
The above equations imply, in particular,
the following simple relation between
$f_1(\rho)$ and $h_2(\rho)$:
\beq
\label{eq:relate_g_h}
h_2(\rho) \, = \, \frac{D_1}{A_1} \, f_1(\rho).
\eeq
We will see an explicit check
of the above equation in the
next section.
However, there is not a similar simple relation
at higher orders, i.e. between $h_3$
and $f_2$, $h_4$ and $f_3$, and so on. 

By replacing the above series
for $f_n(\rho)$ and $h_{n+1}(\rho)$,
eqs.(\ref{eq_fn_Ak}) and (\ref{eq_hnp1_Dk})
respectively,
in the long-distance form factor,
eq. (\ref{eq_def_S_L}),
the latter can
be formally written in terms of a single function series, as follows: 
\bea
S_L &=& \overline{S}_S \, \exp\left\{
\sum_{n=1}^\infty
\alpha_S^{n-1}\left( m^2 \right)
\sum_{k=1}^n
A_k \log\left[ \frac{ \left(c_k \, Q\right)^2 }{m^2} \right]
\, t_k^{(n)}(\rho)
\right\}
\\ \nonumber
&=& \overline{S}_S \, \exp\left\{
A_1 \log\left[ \frac{ \left(c_1 \, Q\right)^2 }{m^2} \right]
\, t_1^{(1)}(\rho)
\, + \,
A_2 \, \alpha_S\left( m^2 \right)
\log\left[ \frac{ \left(c_2 \, Q\right)^2 }{m^2} \right]
\, t_2^{(2)}(\rho) \, + \right.
\\ \nonumber
&& \left.
\qquad\qquad\qquad\qquad\qquad\quad\, 
+ \, A_1 \, \alpha_S\left( m^2 \right)
\log\left[ \frac{ \left(c_1 \, Q\right)^2 }{m^2} \right]
\, t_1^{(2)}(\rho) \, + \, 
\mathcal{O}\left(\alpha_S^2\right)
\right\};
\eea
where we have defined the coefficients
\beq
c_k \, \equiv \, \exp\left(\frac{D_k}{2A_k}\right);
\qquad k = 1,2,3,\cdots.
\eeq
\item
The functions
$f_i(\rho)$ and $h_{i+1}(\rho)$ can be obtained,
for any $i=1,2,3,\cdots$,
from the function $g_{i+1}(\lambda)$
of the usual massless theory,
by looking at the single-logarithmic
terms only, after trivial substitutions
($\lambda \mapsto \rho$, $A_i \mapsto 0$ 
and $B_i \mapsto A_i$ 
or $D_i$, respectively).
\item 
The functions $F\left[ \rho; \alpha_S\left(m^2\right) \right]$ and  $H\left[ \rho; \alpha_S\left(m^2\right) \right]$ do depend, both directly and indirectly through the variable $\rho$, on the QCD coupling
at the scale of the heavy-quark mass, 
$\alpha_S\left(m^2\right)$
(indeed $L \mapsto \mathcal{L}$ and
$\lambda \mapsto \rho$ for $Q^2 \mapsto m^2$).  
Eventually, $\alpha_S\left(m^2\right)$ can be re-expressed, via renormalization group equation, in an equivalent way (modulo ambiguity in the truncation of the QCD running coupling) in
  terms of $\alpha_S\left(Q^2\right)$  and $\log(Q^2/m^2 ) \gg 1$.
\end{enumerate}


\section{Explicit results at Next-to-Next-to-Leading-Logarithmic accuracy}
\label{sec_res_NLL}

In Next-to-Next-to-Leading-Logarithmic accuracy
for both the short-distance form factor and the
long-distance one, namely
\beq
\textrm{N}^2\textrm{LL} + \textrm{N}^2\textrm{LL}_m,
\eeq
in which we work from now on,
the form factors are given respectively by:
\bea
\label{eq_Phi_NLL}
S_S^{\,(\textrm{N}^2\textrm{LL})} &=& \exp\Big\{
L \, g_1(\lambda) \, + \,  g_2(\lambda)
 \, + \, \alpha_S\left(Q^2\right) g_3(\lambda)
\Big\};
\\ \nonumber
S_L^{\,(\textrm{N}^2\textrm{LL}_m)} &=& \overline{S}_S^{\,(\textrm{N}^2\textrm{LL})} \, 
\exp\Bigg\{
\log\left( \frac{Q^2}{m^2} \right)
\Big[ f_1(\rho)
\, + \,
\alpha_S\left(m^2\right) \, f_2(\rho)
\, + \,
\alpha_S^2\left(m^2\right) \, f_3(\rho)
\Big] \, +
\nonumber\\
&& \qquad\qquad\qquad\qquad\quad + \, h_2(\rho) 
\, + \, \alpha_S\left(m^2\right) \, h_3(\rho) 
\Bigg\}.
\eea
By integrating over $k_\perp^2$
all the terms on the r.h.s. of eq.(\ref{eq_main_resumm})
using the usual step-approximation
for the Bessel function
(which is valid up to N$^2$LL or N$^2$LL$_m$ accuracy included),
\beq
1 \, - \, J_0\left(b \, k_\perp \right) \, \simeq \, 
\theta\left( k_\perp^2 \, - \, \frac{b_0^2}{b^2} \right),
\eeq
one obtains for the first three functions in $S_S^{(\textrm{N}^2\textrm{LL})}$:
\bea
\label{eq_give_fi}
g_1(\lambda) &=& 
+ \, \frac{A_1}{\beta_0} 
\, \frac{\lambda \, + \, 
\log(1 \, - \, \lambda)}{\lambda};
\nonumber\\
g_2(\lambda) &=& - \, \frac{A_2}{\beta_0^2} 
\left[
\frac{\lambda}{1-\lambda} \, + \, \log(1-\lambda)
\right]
\, + \, \frac{A_1 \beta_1}{\beta_0^3} \,
\left[
\frac{\log(1-\lambda)+\lambda}{1-\lambda}
\, + \, \frac{1}{2} \log^2(1-\lambda)
\right] \, +
\nonumber\\
&& + \, \frac{B_1}{\beta_0} \log(1-\lambda) ;
\nonumber\\
g_3(\lambda) &=& 
- \, \frac{A_3 }{2 \beta_0^2} \, \frac{\lambda^2}{(1-\lambda)^2}
\, - \, \frac{{B}_2}{\beta_0} \, \frac{ \lambda}{ 1-\lambda}
\, - \,  \frac{A_2 \, \beta_1}{2  \beta_0^3} \,  
\frac{ \lambda (2-3 \lambda) \, + \, 2 (1-2 \lambda) \, \ln(1-\lambda)}{(1-\lambda)^2}  \, +
\nonumber\\   
&&  + \, \frac{A_1 \, \beta_2 }{2  \beta_0^3} \, 
\left[
 \frac{\lambda (2-3 \lambda)}{(1-\lambda)^2} \, + \, 2 \ln(1-\lambda) \right]  
\, + \, \frac{{B}_1 \, \beta_1}{\beta_0^2} \, 
\frac{ \lambda \, + \, \ln (1-\lambda) }{ 1-\lambda} \, +       
\nonumber\\
&& + \, \frac{A_1 \, \beta_1^2}{2 \, \beta_0^4} \, 
\frac{ \lambda^2 \, + \, (1-2\lambda) \ln^2(1-\lambda) 
\, + \, 2 \lambda (1-\lambda) \, \ln (1-\lambda) }{ (1-\lambda)^2}. 
\eea
As far as the function-series expansion of $F(\rho;\alpha_S)$
is concerned, the first three terms explicitly read:
\bea
\label{eq_define_gs}
f_1(\rho) &=& + \, \frac{A_1}{\beta_0} \log(1-\rho) ;
\nonumber\\
f_2(\rho) &=& - \, \frac{A_2}{\beta_0} 
\, \frac{\rho}{1-\rho}
\, + \, \frac{A_1 \beta_1}{\beta_0^2} \,
\frac{\log(1-\rho) + \rho}{1-\rho};
\nonumber\\
f_3(\rho) &=&
- \, \frac{A_3}{2 \, \beta_0} \, 
\frac{\rho \, (2 \, - \, \rho)}{(1 \, - \, \rho)^2}
\, + \, \frac{A_2 \, \beta_1}{2 \, \beta_0^2}
\, \frac{2 \log(1 \, - \, \rho) 
\, + \, \rho \, (2 \, - \, \rho) }{(1 \, - \, \rho)^2}
\, - \, \frac{A_1 \, \beta_2}{2 \, \beta_0^2}
\frac{\rho^2}{(1 \, - \, \rho)^2} \, +
\nonumber\\
&& \,\, + \, \frac{A_1 \, \beta_1^2}{2 \, \beta_0^3}
\, \frac{\rho^2 \, - \, \log^2(1 \, - \, \rho)}{(1 \, - \, \rho)^2}.
\eea
The first two non-vanishing terms in the function series for $H(\rho,\alpha_S)$
are written ($h_1(\rho)\equiv 0$):
\bea
\label{eq_define_g2}
h_2(\rho) &=&  + \, \frac{D_1}{\beta_0} \, \log(1 - \rho);
\nonumber\\
h_3(\rho) &=& - \, \frac{D_2}{\beta_0}
\, \frac{\rho}{1-\rho}
\, + \, \frac{D_1 \, \beta_1}{\beta_0^2}
\, \frac{\log(1-\rho) \, + \, \rho}{1-\rho}.
\eea
The first equation above is in complete agreement with eq.(\ref{eq:relate_g_h}).

\noindent
The explicit expressions of the first order coefficients of the functions $A(\alpha_S)$, $B(\alpha_S)$ and $D(\alpha_S)$ are given in eq.(\ref{abd1}) while the higher order coefficients
read:
\bea
A_2 &=& \frac{C_F}{\pi^2}
\left[
C_A \left( \frac{67}{36} \, - \, \frac{\pi^2}{12} \right)
\, - \, \frac{5}{18} n_f
\right];
\nonumber\\
A_3 &=& \frac{C_F}{\pi^3}
\Bigg[ \,
C_A^2 \left(
\frac{15503}{2592} - \frac{67}{36} z_2 
+ \frac{11}{8} z_4 - \frac{11}{4} z_3
\right)
+ C_A n_f \left(
- \frac{2051}{1296}
+ \frac{5}{18} z_2
\right) \, +
\nonumber\\
&&\quad  \,\, + \, C_F \, n_f \left( 
- \frac{55}{96} + \frac{z_3}{2}
\right)
\, + \, \frac{25}{324} n_f^2
\Bigg];\nonumber
%
\\ \nonumber
B_2 &=& + \, \frac{C_F}{\pi^2}
\Bigg[
C_A \left(
- \frac{35}{16} + \frac{11}{4} z_2 + \frac{3}{2} z_3
\right)
\, + \, 
C_F\left( - \frac{3}{16} + \frac{3}{2} z_2 - 3 z_3  \right)
\, + \, 
n_f \left( \frac{3}{8} - \frac{z_2}{2}\right)
\Bigg].
\eea
As usual, $C_A=N_C=3$ in QCD and $n_f$ is the number of active 
(i.e. effectively massless) quark flavors.
The constant $z_n$ is the value of the Riemann Zeta function
$Z(x)$ at the integer point $x=n$.

The value of the coefficient $D_2$ is presently unknown. We assume, rather heuristically, that the coefficient $D_2$ is exactly given by the
two-loop contribution to the cusp anomalous dimension \cite{Korchemsky:1987wg,Aglietti:2008xn}\footnote{We take the opportunity to point out a
typographical mistake in Eq.(9) of Ref.\cite{Aglietti:2008xn}.
The term $-\frac{5}{18}\,n_f$ should actually read $+\frac{5}{9}\,n_f$.}:
\bea
\label{d2}
D_2 &=& + \, \frac{C_F}{2\pi^2}
\left[
C_A \left(
z_2 \, - \, z_3 \, - \, \frac{49}{18}
\right)
\, + \, \frac{5}{9} \, n_f
\right].
\eea
Therefore, given the above assumption about the size of the $D_2$ term,
the inclusion of the NNLL corrections is not as rigorous as in the NLL case.
An explicit check of this assumption can be obtained by comparing the
expansion of our resummation formula with an explicit (massive) two-loop
computation of the EEC function, which is currently not available.

For completeness sake, 
let us also give the explicit expressions of the first three coefficients of the 
QCD $\beta$-function with our normalization
(involving just powers of $\alpha_S$, without $\pi$
at the denominator, $d\alpha_S/d \log\mu^2 = - \beta_0 \, \alpha_S^2 -\beta_1 \, \alpha_S^3-\beta_2 \, \alpha_S^4 +\cdots$):
\bea
\beta_0 &=& \frac{1}{4\pi} 
\left( \frac{11}{3}C_A - \frac{2}{3} n_f \right)
\, = \, \frac{33 \, - \, 2 \, n_f}{12 \pi};
\nonumber\\
\beta_1 &=& \frac{1}{24\pi^2} 
\big( 17 \, C_A^2 - 5 \, C_A n_f - 3 \, C_F n_f \big)
\, = \, \frac{153 \, - \, 19 \, n_f}{24 \pi^2};\nonumber
\\ \nonumber
\beta_2 &=&  \frac{1}{64\pi^3} 
\left[
\frac{2857}{54} \, C_A^3
\, - \, 
\left(
\frac{1415}{54} C_A^2 \, + \, \frac{205}{18} C_A C_F \, - \, C_F^2
\right) n_f
\, + \,
\left(
\frac{79}{54} C_A \, + \, \frac{11}{9} C_F 
\right) n_f^2
\right] \, =
\nonumber\\
&=& 
\frac{1}{128 \, \pi^3}
\left(
2857 \, - \, \frac{5033}{9} \, n_f \, + \, \frac{325}{27} \, n_f^2
\right).
\eea
The following remarks are in order.
\begin{enumerate}
\item
As already noted in the general case
(see previous section), 
the functions $g_i(\lambda)$ are 
exactly equal to the corresponding ones
of the massless case.
All these functions have a singularity,
related to the Landau Pole (LP) in $\alpha_S(k_\perp^2)$,
at the point
\beq
\lambda \, = \, 1,
\eeq
i.e. for a value of the impact parameter
\beq
b_{\mathrm{LP}} \, \equiv \, \frac{b_0}{Q} \, 
\exp\left[ \frac{1}{2 \, \beta_0 \, \alpha_S\left(Q^2\right)} \right]
.
\eeq
In the case of a heavy quark, i.e. a quark with a mass
\beq
m \, \gg \, \Lambda_{QCD},
\eeq
it holds:
\beq
b_{\mathrm{CR}} \, \ll \, b_{\mathrm{LP}},
\eeq
so that in $S_S(b)$ the Landau singularity
does not show up.
On the contrary, in the case of a light quark, the Landau singularity
does occur in $S_S(b)$;
All that simply implies that mass effects, in the case
of light quarks ($m=\mathcal{O}(\Lambda_{QCD})$),
cannot be described by resummed perturbation theory.
These facts are in complete agreement with one's physical
intuition.
\item
The functions $F(\rho;\alpha_S)$ and $H(\rho;\alpha_S)$
do have a singularity related to the Landau pole
at the point $\rho=1$, i.e. at a value of the $b$-parameter
as large as
\beq
\tilde{b}_{\mathrm{LP}} \, \equiv \, \frac{b_0}{m} \, 
\exp\left[ \frac{1}{2 \, \beta_0 \, \alpha_S\left(m^2\right)} \right]
\, \simeq \, b_{\mathrm{LP}}.
\eeq
The last equality is a consequence of the $\beta$-function flow.
Since the long-distance form factor is defined for all distances $b>b_{\mathrm{CR}}$, and 
$b_{\mathrm{LP}} > b_{\mathrm{CR}}$ for a heavy quark, 
$S_L(b)$ is effectively affected by Landau-pole ambiguities.

If nonperturbative effects do occur {\it for all} $b>b_{\mathrm{CR}}$,
as it could happen for example in the case of the charm quark,
one may think of replacing 
the usual QCD coupling with its analytic
space-like or time-like regularizations 
\cite{Aglietti:2006yb,Aglietti:2006yf}.
If nonperturbative effects are found to be very large,
one can still assume, 
for the exponent of $S_L$,
the perturbative structure
linear in $\log(Q^2/m^2)$,
but replace
the perturbative functions $F(\rho;\alpha_S)$ and $H(\rho;\alpha_S)$
by two (unknown) nonperturbative functions, 
so that:
\beq
S_L \,\, \mapsto \,\, \overline{S}_S \, 
\exp\left\{
\log\left( \frac{Q^2}{m^2} \right) 
\, F_{\mathrm{NP}}\left(b\right)
\, + \, H_{\mathrm{NP}}\left(b\right)
\right\}.
\eeq
The two above functions, which we have written
directly as functions of the impact parameter $b$, 
have to be determined experimentally,
by fitting EEC spectra at least at two (or more) different values of $Q$.
\item
As already noted in general in the previous section,
the subleading, soft $D$-terms produce
"rescaling effects"
on the hard scale $Q$ (or on the dead-cone
(DC) opening angle $\theta_{DC}$).
For clarity's sake,
let us consider first the simpler NLL$_m$ case
and then the more elaborated N$^2$LL$_m$ case:
\begin{enumerate}
\item[i)]
NLL$_m$ {\it case}.

We have found that the functions $f_1(\rho)$ 
(the first of eqs.(\ref{eq_define_gs}))
and $h_2(\rho)$ (eq.(\ref{eq_define_g2}))
have the same functional form, so one can combine 
the corresponding contributions to $S_L$ into a single term
as follows:
\beq
\log\left(\frac{Q^2}{m^2}\right) f_1(\rho) \, + \, h_2(\rho)
\, = \, \log\left[\frac{(c_1 Q)^2}{m^2}\right] f_1(\rho);
\eeq
where the first-order coefficient explicitly
reads:
\beq
c_1 \,\equiv\, 
\exp\left( \frac{D_1}{2 \, A_1} \right) \, = \, e^{\, - \, 1/2}
\, \simeq \, 0.606531.
\eeq
We can interpret the above equation
by saying that the subleading function $h_2(\rho)$
has the {\it sole effect} of reducing
by a factor $1/c_1 \simeq 1.64872$
the hard scale $Q$ entering the leading 
$f_1(\rho)$ term%
\footnote{
As well known, one can play a similar trick also in the
short-distance form factor by writing,
for example, at first order:
\beq
A_1 \log\left(\frac{Q^2}{k_\perp^2}\right) \,+ \, B_1
\,\,\, \mapsto \,\,\,
A_1 \log\left[ \frac{Q^2 \exp(B_1/A_1)}{k_\perp^2} \right].
\eeq
However, in this case, the hard scale squared $Q^2$ also appears
in the upper integration limit on $k_\perp^2$,
so an interpretation of the $B_1$ term
similar to that of the $D_1$ 
is not possible.
}.
%
The classical --- or Leading-Log massive (LL$_m$) --- 
dead-cone restriction reads:
\beq
\theta \,\, \gsim \,\, 
\theta_{DC}^{(LL_m)} \, \equiv \, \frac{2m}{Q},
\eeq
where we have taken into account that the quark energy $E \simeq Q/2$ in the soft region.
The $D_1$-effect then implies
\beq
\theta \,\, \gsim \,\, \theta_{DC}^{(NLL_m)}
\, \equiv \,
\frac{\theta_{DC}^{(LL_m)}}{c_1}. 
\eeq
Therefore the subleading soft $D_1$ effect increases
the dead-cone opening angle by the factor
$1/c_1 \simeq 1.64872$
\footnote{
The related problem of finding observables
with good sensitivity with respect to the
(perturbative) dead-cone effect
has recently been treated in \cite{Dhani:2024gtx}.}.
\item[ii)]
N$^2$LL$_m$ {\it case}.

The combination of the terms proportional to
$f_2(\rho)$ and $h_3(\rho)$ in $S_L$
is substantially more
complicated than in the previous NLL$_m$ case,
basically because of subleading running-coupling 
($\propto \beta_1$) effects:
\bea
\label{eq_two_rescalings}
&&\alpha_S\left(m^2\right) 
\left[
\log\left(\frac{Q^2}{m^2}\right) f_2(\rho) \, + \, h_3(\rho)
\right] \, =
\\ \nonumber
&=& 
\log\left[ \frac{ \left(c_1 \, Q\right)^2}{m^2} \right]
\frac{A_1 \, \beta_1}{\beta_0^2} \,
\frac{\log(1 \, - \, \rho) \, + \, \rho}{1\, -\, \rho}
\, - \, \log\left[\frac{ \left(c_2 \, Q\right)^2 }{m^2} \right]
\frac{A_2}{\beta_0} \, \frac{\rho}{1-\rho} .
\eea
The second-order coefficient explicitly reads:
\bea
c_2 &\equiv& \exp\left(\frac{D_2}{2A_2}\right) \, = \,
\exp \left[ - \, \frac{C_A \big(
49/18 \, - \, z_2 \, + \, z_3  
\big)
\, - \, 5/9 \, n_f}{
C_A \big( 67/9 \, - \, 2 \, z_2 \big)
\, - \, 10/9 \, n_f}
\right].
\eea
In the above ratio,
the dependence on the number of active flavors
largely cancels, so the variation
of $c_2 = c_2\left(n_f\right)$ with $n_f$
is rather mild. 
Indeed, in the purely-gluonic theory:
\beq
c_2(n_f=0) \, \simeq \, 0.577738;
\eeq
while, for five active flavors ($Q>m_b$):
\beq
c_2(n_f=5) \, \simeq \, 0.555578.
\eeq
Therefore, for $n_f:0\mapsto 5$, there is  only a $4\%$ reduction of $c_2(n_f)$.
Note also that $c_2(n_f=5)$ is less than $10\%$ smaller than $c_1$,
so its hard-scale reducing effect is only slightly larger.

By looking at the r.h.s. of eq.(\ref{eq_two_rescalings}), we find that, beyond NLL$_m$, it is not
anymore possible to describe the 
soft $D$-type effects in terms of a
{\it unique} rescaling of the hard scale $Q$,
or of the dead-cone opening angle $\theta_{DC}$.
Indeed, at the N$^2$LL$_m$ level,
in order to describe the effects of the
soft function $h_3(\rho)$, 
it is rather necessary to introduce {\it two different rescalings} of $Q$ or $\theta_{DC}$: one of them
--- related to the $A_1 \beta_1$ term --- involving the old, NLL$_m$
coefficient $c_1$; the other one
--- related to the $A_2$ term ---
involving the new coefficient $c_2$.

\end{enumerate}
\item
At first order in $\alpha_S$, the $b$-space Sudakov form factor
explicitly reads:
\bea
S^{(1)}(b) &=& \theta\left( b_{\mathrm{CR}} - b \right)
\, \exp\left\{ - \, \frac{A_1}{2}\alpha_S \log^2\left( \frac{Q^2 b^2}{b_0^2} \right)
\, - \, B_1 \alpha_S \log\left( \frac{Q^2 b^2}{b_0^2} \right)
\right\} \, +
\nonumber\\
&+& \theta\left( b - b_{\mathrm{CR}} \right)
\, \exp\Bigg\{ 
- \, \frac{A_1}{2}\alpha_S \log\left(\frac{Q^2}{m^2}\right)
\left[
\log\left(\frac{Q^2}{m^2}\right)
\, + \, 2\log\left(\frac{m^2 b^2}{b_0^2}\right)
\right] \, +
\nonumber\\
&& \qquad\qquad\qquad\quad\,\,\,\,
- \, B_1 \alpha_S \log\left(\frac{Q^2}{m^2}\right)
\, - \, D_1 \alpha_S \log\left(\frac{m^2 b^2}{b_0^2}\right)
\Bigg\}.
\eea
\item
If we consider a process with a hard scale
$Q > m_b$, as for example $e^+e^-$ annihilation 
to hadrons at the $Z^0$ peak, 
one has to consistently set $n_f=5$ in the
short-distance form factor and  
$n_f=4$ in the long-distance one.
That is because $S_S$ only involves $k_\perp > m_b$
transverse momenta,
while $S_L$ only involves  $k_\perp < m_b$ transverse momenta.
In practice, one sets $n_f=5$ in 
the coefficients entering the r.h.s. of eqs.(\ref{eq_give_fi})
and in the $\beta_0$-coefficient
entering the definition of $\lambda$,
first of eqs.(\ref{eq_def_lambda_and_rho}); 
one sets $n_f=4$ in eqs.(\ref{eq_define_gs})
and (\ref{eq_define_g2}),
and in the $\beta_0$-coefficient
entering the definition of $\rho$,
second of eqs.(\ref{eq_def_lambda_and_rho}).
\end{enumerate}


\section{Numerical Results}
\label{sec_phenomen}

\begin{figure}[th]
\begin{center}
\includegraphics[width=.5\textwidth]{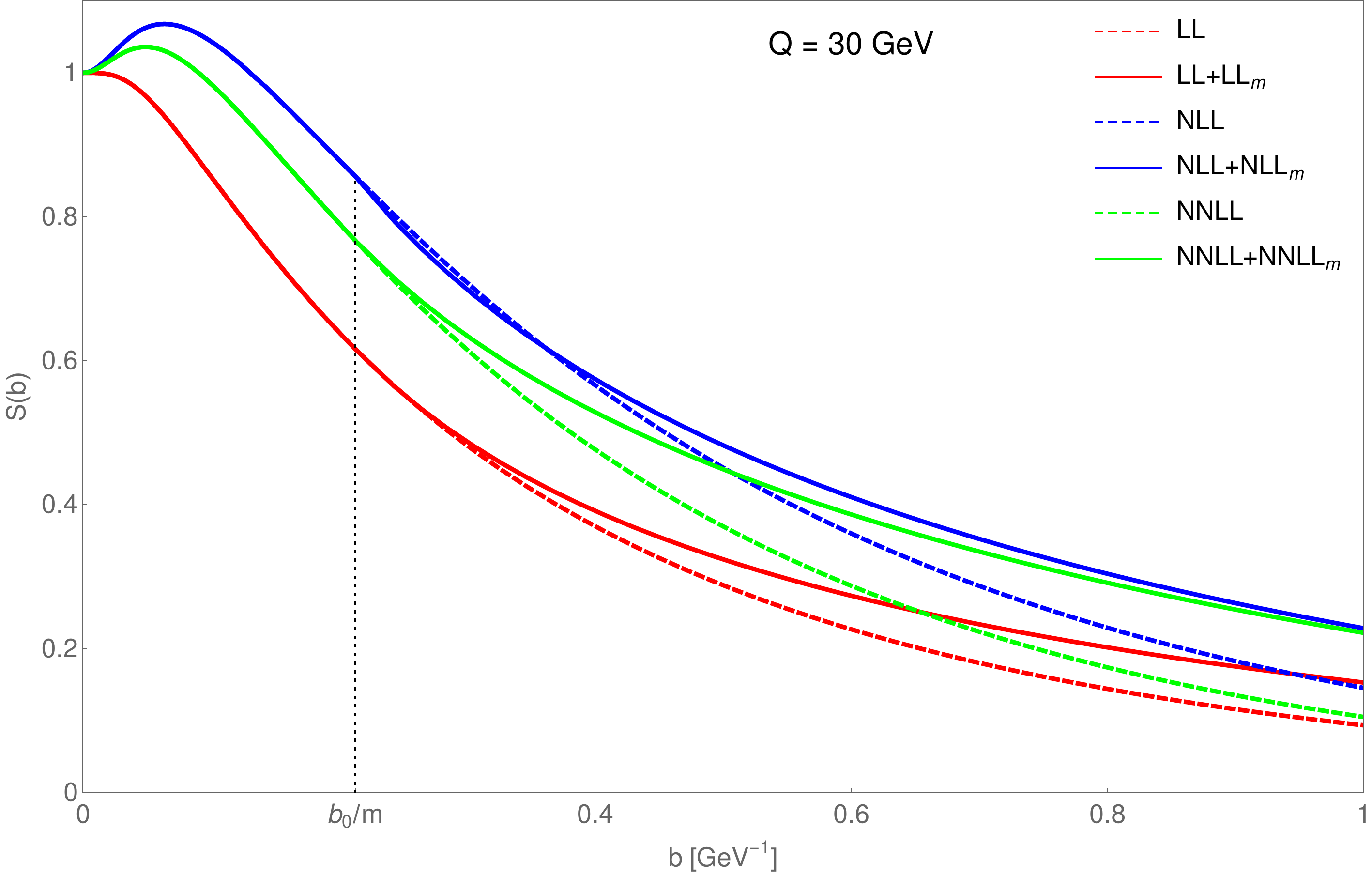}
\includegraphics[width=.505\textwidth]{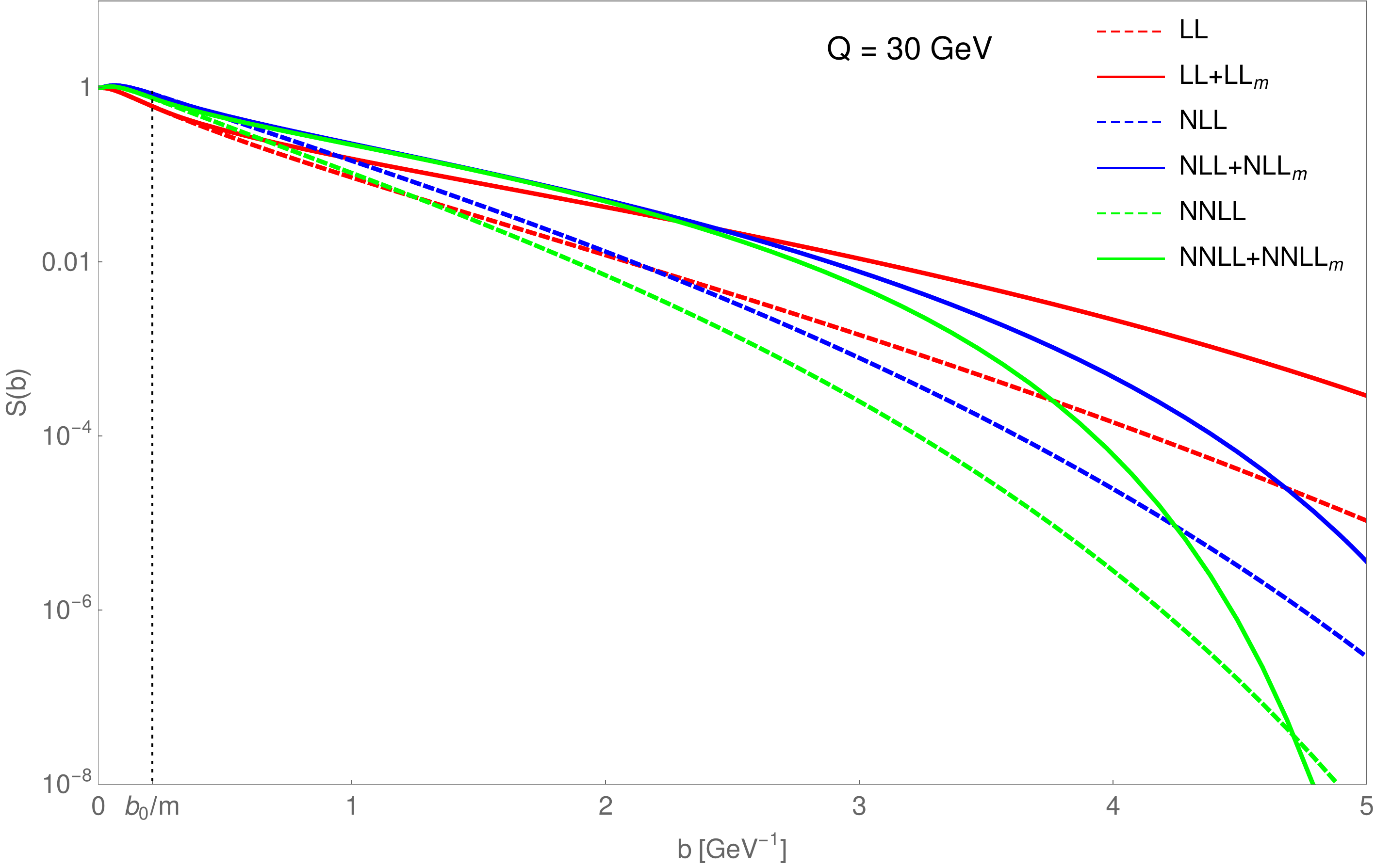}
\end{center}
\caption{
\label{figb1}
{\em
The Sudakov form factor $S(b)$ in $b$-space at a hard scale $Q=30$~GeV both in linear (left panel) and logarithmic (right panel) scales on the vertical axis, at various logarithmic orders.
The solid lines represent the massive case with $m=m_b$, while the dashed lines the massless case.
}}
\end{figure}

\begin{figure}[th]
\begin{center}
\includegraphics[width=.5\textwidth]{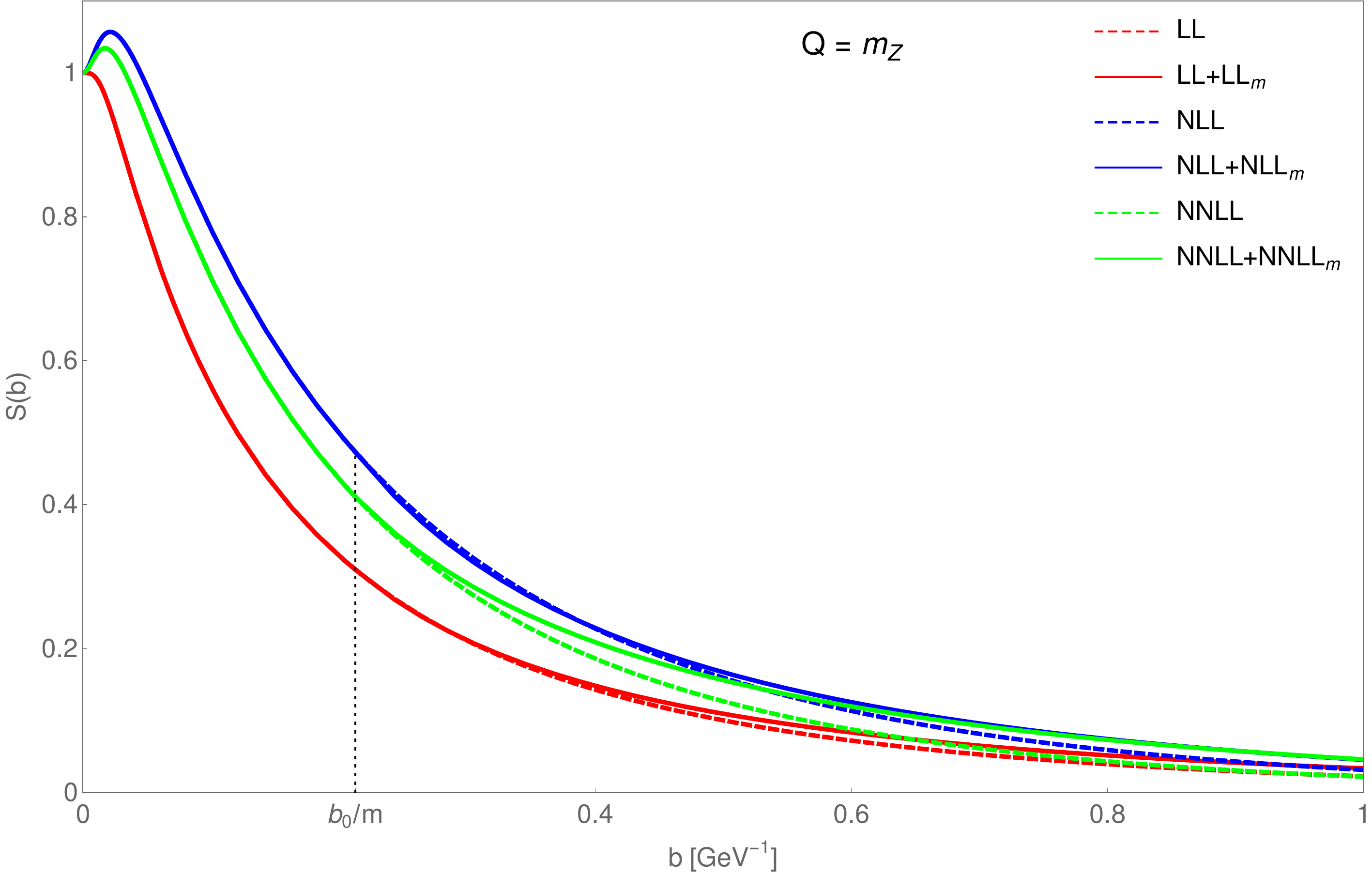}
\includegraphics[width=.505\textwidth]{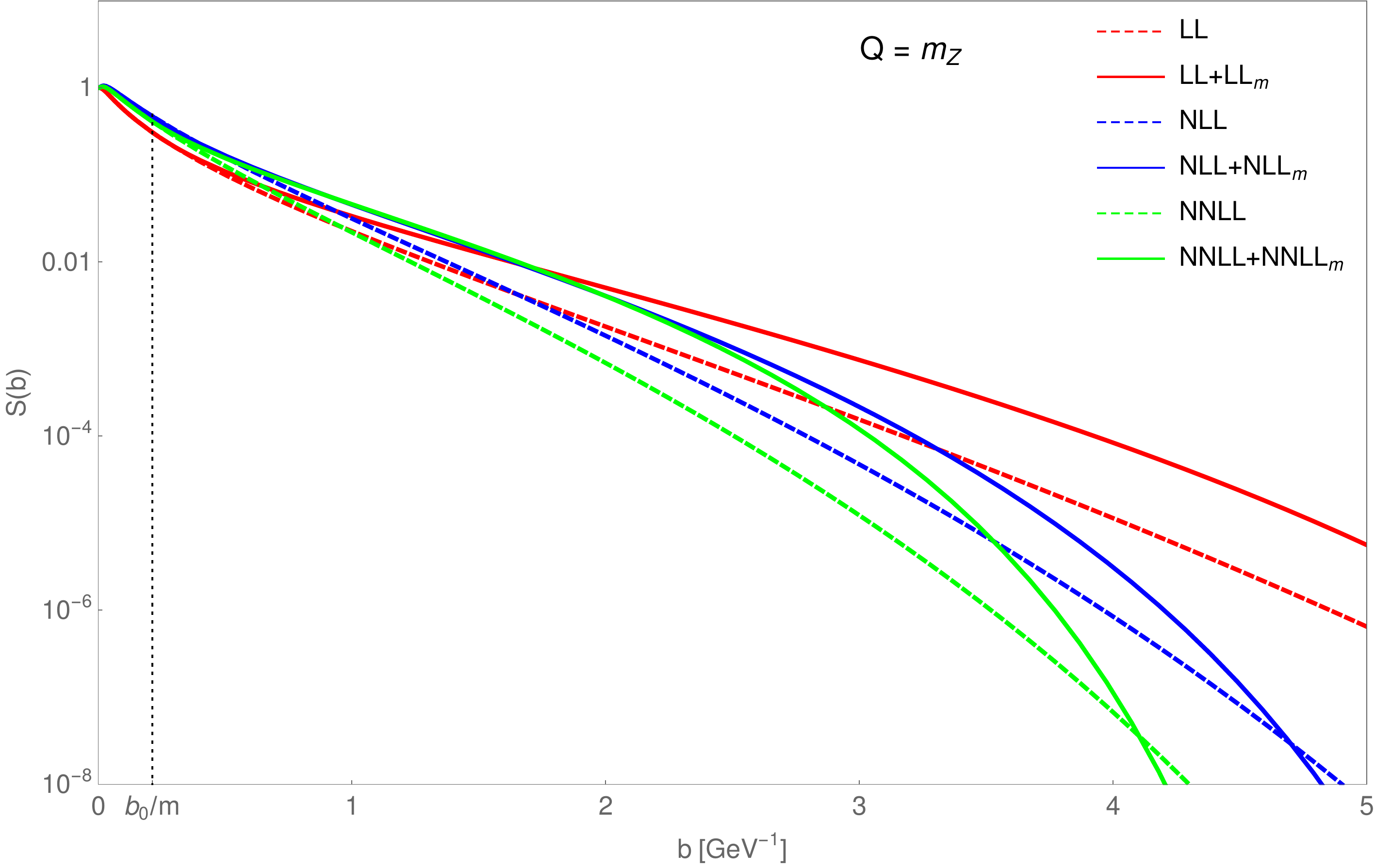}
\end{center}
\caption{
\label{figb2}
{\em
The Sudakov form factor $S(b)$ in $b$-space at the $Z^0$ peak, $Q=m_Z$, both in linear (left panel) and logarithmic (right panel) scales on the vertical axis.
The solid lines are the massive case with $m=m_b$, while the dashed lines are the massless case.
}}
\end{figure}

\noindent
In this section we analyze the
impact of heavy quark mass corrections
to EEC spectra, which 
we have derived in the previous section.
%
%
In fig.\ref{figb1}, we plot the Sudakov
form factor in $b$-space at LL+LL$_m$ (red solid line), NLL+NLL$_m$ (blue solid line) and NLL+NLL$_m$ (green solid line) accuracy,
for a hard scale $Q=30\,\mathrm{GeV}$,
for beauty quarks with a mass $m_b=m_B=5.278\,\mathrm{GeV}$%
\footnote{
We {\it define} the beauty mass $m_b$
as the mass of the lowest-lying hadron
containing a beauty quark, 
namely a $B$ meson,
so that $m_b=m_B$.
Other choices are, of course, possible.
},
together with the corresponding massless case (dashed lines). 
As expected by the suppression
of gluon radiation produced by the dead-cone
effect,
the massive form factor
decays
slower to zero 
in the long-distance region
$(b>b_{\mathrm{CR}})$
than the massless form factor.

%
%
In fig.\,\ref{figb2},
in order 
to estimate the effects of varying the hard scale $Q$, we plot the Sudakov
form factor in $b$-space at
LL+LL$_m$ (red solid line), NLL+NLL$_m$ (blue solid line) and NLL+NLL$_m$ (green solid line) accuracy
at the $Z^0$ peak, i.e. for a hard scale $Q=m_Z \simeq 91.19\,\mathrm{GeV}$,
again for beauty quarks with $m_b=m_B$
and in the corresponding massless case (dashed lines). 
As expected,
by increasing the hard scale by a factor
three, mass effects are roughly reduced by
an order of magnitude.

\begin{figure}[th]
\begin{center}
\includegraphics[width=.75\textwidth]{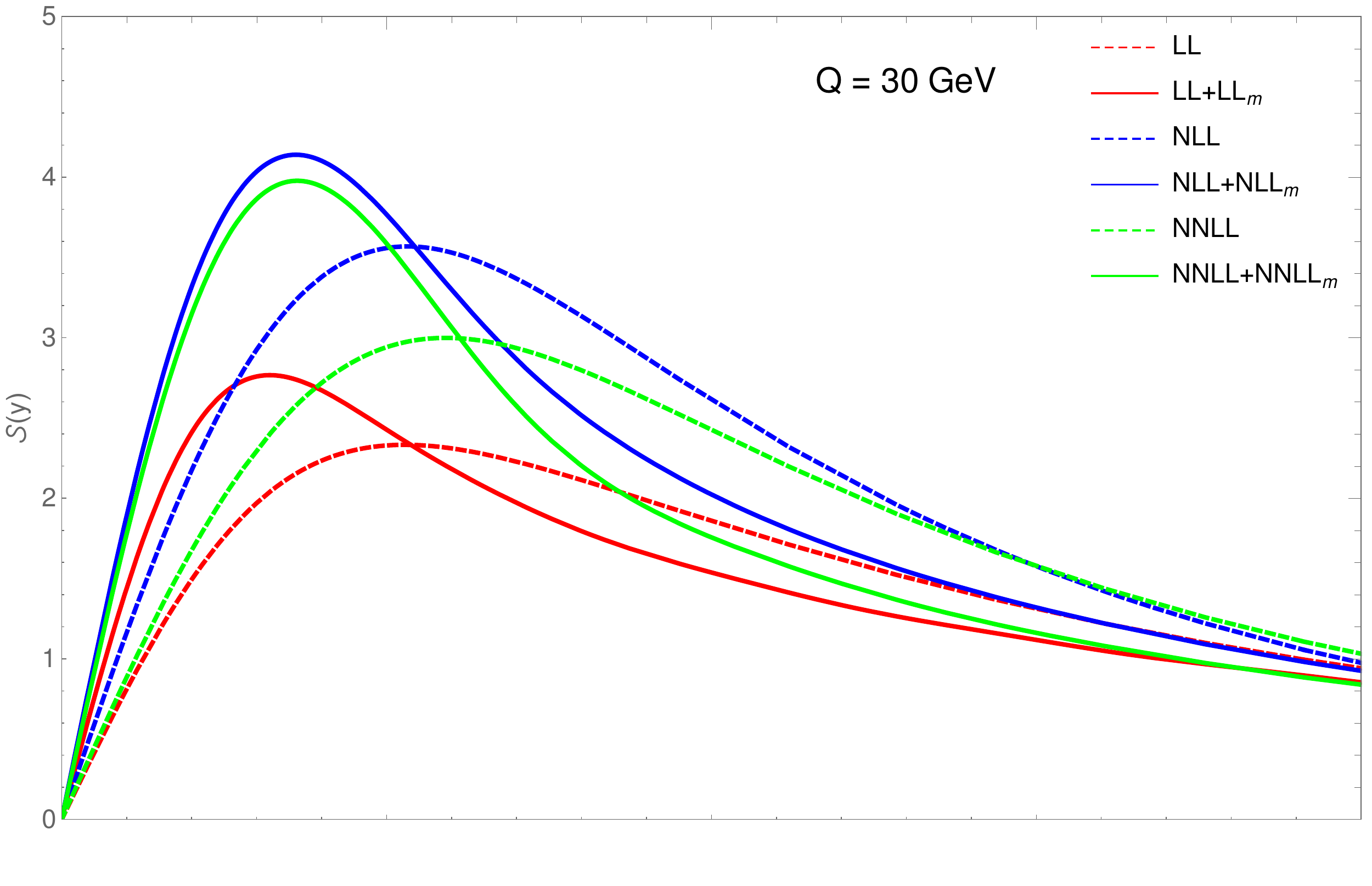}\vspace*{-.5cm}\\
\hspace*{-.1cm}\includegraphics[width=.77\textwidth]{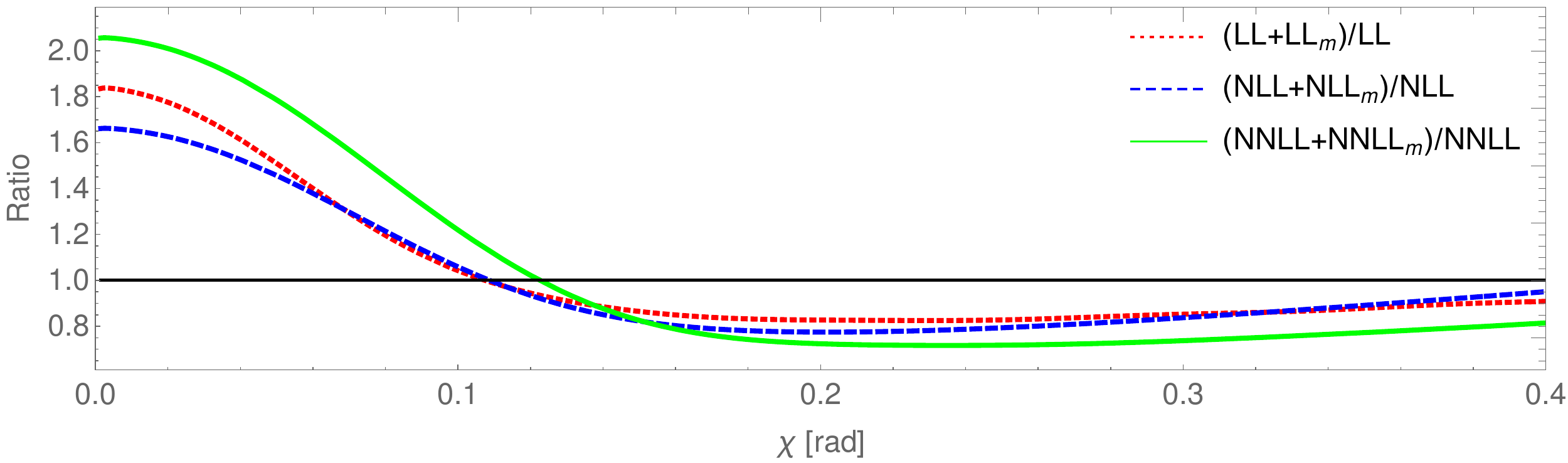}
\end{center}
\caption{
\label{figchi1}
{\em
The Sudakov form factor ${\cal S}(y)$ in physical (angle-$\chi$) space for a hard scale $Q=30$~GeV.
The solid lines are the massive case with $m=m_b$, while the dashed lines are the massless case.
In the lower panel it is shown the ratio of the massive form factor over the massless one.
}}
\end{figure}

\begin{figure}[th]
\begin{center}
\includegraphics[width=.75\textwidth]{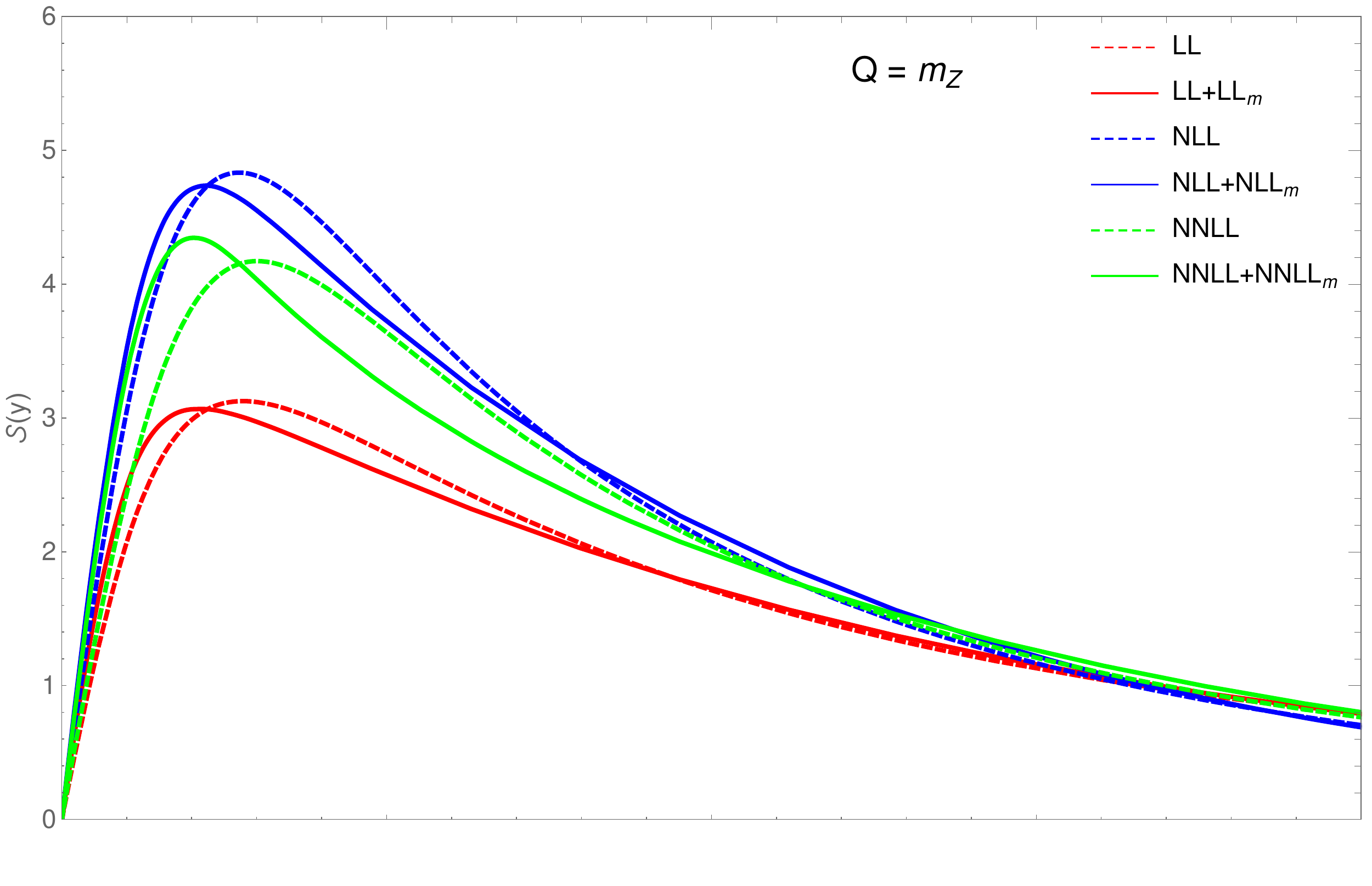}\vspace*{-.5cm}\\
\hspace*{-.1cm}\includegraphics[width=.77\textwidth]{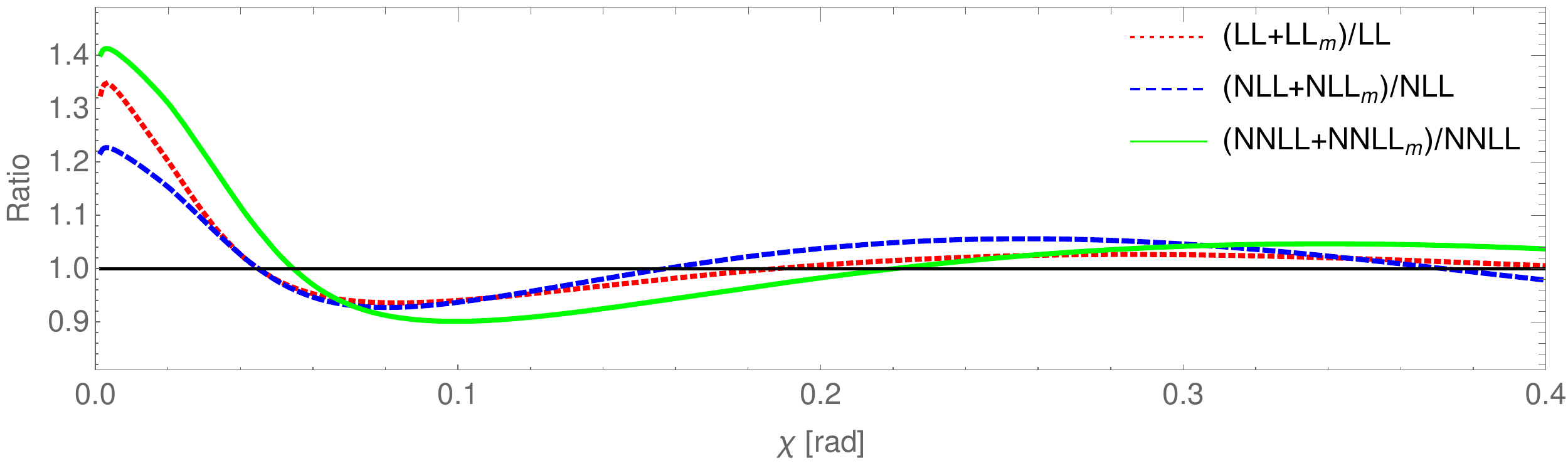}
\end{center}
\caption{
\label{figchi2}
{\em        
The Sudakov form factor ${\cal S}(y)$ in physical (angle-$\chi$)  space at the $Z^0$ peak, i.e. for $Q=m_Z$.
The solid lines represent the massive case with $m=m_b$, while the dashed lines describe the massless case.
In the lower panel it is shown the ratio massive/massless form factor.
}}
\end{figure}

In fig.\,\ref{figchi1} we plot the Sudakov form factor
${\cal S}(y)$ in the physical space of the 
correlation angle $\chi$, 
for $Q=30\,\mathrm{GeV}$ in the beauty case
at LL+LL$_m$ (red solid line), NLL+NLL$_m$ (blue solid line) and NLL+NLL$_m$ (green solid line) accuracy
and in the massless case (dashed lines).
We observe that
massive distributions have a peak which is higher, thinner
and closer to the back-to-back point 
$\chi=0$ than the peak of the
corresponding massless distributions.
These are all clear consequences of the suppression
of gluon radiation by the dead cone effect.

If one does not "tag", i.e. does not identify
the $b\bar{b}$ final states in the process (\ref{reaction_b-only}) and only
measures the EEC distribution in the flavor-inclusive
case,
\beq
e^+e^- \, \to \, q\bar{q} \, + \, \cdots,
\eeq
with $q=u,d,s,c,b$, then the beauty-mass effects,
for $Q \ll m_Z$,
by neglecting the $Z^0$ exchange diagram,
are roughly reduced by a factor
\beq
\frac{e_b^2}{\sum_{f=1}^5
e_f^2} \, = \, \frac{1}{11}\,,
\eeq
where $e_f$ is the electric charge 
of quark flavor $f$.
A reduction of the mass effects
shown in fig.\,\ref{figchi1}
basically by an order of magnitude is then
predicted.

%
%
In fig.\,\ref{figchi2} we plot the Sudakov form factor of  ${\cal S}(y)$ in the physical space
at the $Z^0$ peak in the beauty case
at LL+LL$_m$ (red solid line), NLL+NLL$_m$ (blue solid line) and NLL+NLL$_m$ (green solid line) accuracy
and in the massless case (dashed lines).
Compared to the previous case,
mass effects are clearly much smaller.

If one, at the $Z^0$ peak, is inclusive in quark flavor $f$, 
then by neglecting to a first approximation the photon exchange diagram, 
the beauty mass effects 
shown in fig.\,\ref{figchi2}
are reduced by a factor \cite{Ellis:1996mzs}
\beq
\frac{V^2(d) \, + \, A^2(d)}{
2\left[V^2(u) \, + \, A^2(u)\right]
\, + \, 3\left[V^2(d) \, + \, A^2(d) \right] }
\, \simeq \, 0.219\,,
\eeq
where $V(f)$ and $A(f)$ are the polar-vector and 
axial-vector couplings of flavor $f$ to the $Z^0$ respectively,
\bea
V(f) &=& T_3(f) - 2 \, Q(f) \sin^2(\theta_W);
\nonumber\\
A(f) &=& T_3(f)
\eea
for $f=u,d$ (an {\it up}-type or a {\it down}-type quark).
$Q(f)$ is the electric charge 
and $T_3(f)$ is the third component 
of the weak isospin of flavor $f$.
For the Weinberg angle,
we have taken the numerical value
\cite{ParticleDataGroup:2020ssz}:
\beq
\sin^2(\theta_W) \, \simeq \, 0.2234.
\eeq
Therefore, for the EEC at the $Z^0$ peak
in the flavor-inclusive case, 
beauty quark mass effects are reduced
by a factor $\simeq 4.5$,
obtaining typical 
effects at few percent level.


\section{Conclusions}
\label{sec_concl}

In this work we have analyzed the effects of heavy quark 
masses on the EEC function in $e^+e^-$ annihilation
to hadrons at high energy, in the back-to-back (or two jet) region.
The main results are provided
by eq.(\ref{eq_main_resumm}), eq.(\ref{eq_def_Phi}) and 
by eqs.(\ref{eq_Phi_NLL}).
In general, the formulae we have obtained, turn out to be quite simple and physically very intuitive.

\noindent
In eq.(\ref{eq_main_resumm}), we have presented a general QCD
resummation formula for the EEC Sudakov form factor 
in impact-parameter ($b$-)space for massive quarks
(with mass $m \ll Q$).
We can say that the heavy quark mass $m$ has a "discriminating power"
as far as soft-gluon transverse momenta 
$k_\perp$ are concerned.
Therefore the resummation formula involves 
basically two different
contributions.
The first contribution, of short-distance character,
involves gluon transverse momenta larger than
the heavy quark mass,
\beq
k_\perp^2 \, > \, m^2,
\eeq 
and has the same structure (integrand) 
of the well-known massless resummation formula:
\beq
\hat{S}(b) 
\, = \, \exp \, - \int\limits_{m^2}^{Q^2} 
\frac{dk_\perp^2}{k_\perp^2}
\left\{
\log\left(\frac{Q^2}{k_\perp^2}\right) \,
A\left[\alpha_S\left(k_\perp^2\right)\right]
\, + \, 
B\left[\alpha_S\left(k_\perp^2\right)\right]
\right\}
\Big[
1 \, - \, J_0\left(b k_\perp\right)
\Big]. 
\eeq
The second contribution, of long-distance character,
involves instead gluon transverse momenta smaller than
the heavy quark mass,
\beq
k_\perp^2 \, < \, m^2,
\eeq
and can be written in the suggestive form:
\beq
\label{eq_DeltaS_concl}
\Delta S(b) 
\, = \, \exp \, - \int\limits_0^{m^2} 
\frac{dk_\perp^2}{k_\perp^2}
\left\{
\sum_{n=1}^\infty
A_n \log\left[ 
\frac{Q^2 \exp\left( D_n/A_n \right) }{m^2} \right] 
\, \alpha_S^n\left(k_\perp^2\right)
\right\}
\Big[
1 \, - \, J_0\left(b k_\perp\right)
\Big];
\eeq
where $A_n$ and $D_n$ are coefficients
which can be computed by means of 
fixed-order perturbation theory
(Feynman diagrams).
In order to obtain a consistent theory,
one has then to insert in place of
$\alpha_S\left(k_\perp^2\right)$
its asymptotic expansion.
Note that, in the above formula, there is at most a single
infrared logarithm  
coming from the $k_\perp$
integration, for each power of $\alpha_S$. 
It is interesting to observe that,
by simply imposing the dead-cone restriction
to the massless resummation expression,
one correctly reproduces all the terms
in the massive resummation formula,
with the exception of the $D_n$ terms.

In eq.(\ref{eq_def_Phi}), we have given the general form
of the usual function-series representation
of the Sudakov form factor, which is at the basis of the 
well-known fixed-logarithmic
accuracy.
The form factor has a different form 
at small distances, 
\beq
b \, \lsim \, \frac{1}{m},
\eeq
where it is equal to the
massless form factor, and at large distances,
\beq
b \, \gsim \, \frac{1}{m},
\eeq
where mass effects are substantial and
are described by two different functions series. 
It turns out that one can introduce
{\it independent} logarithmic approximation schemes
for the short-distance form factor 
and the long-distance one.
For the short-distance form factor, $S_S$,
one has the usual logarithmic-accuracy scheme: 
Leading-Log 
(LL) approximation (keep the function $g_1(\lambda)$ only); 
Next-to-Leading-Log
(NLL) approximation (keep also $g_2(\lambda)$); 
Next-to-Next-to-Leading-Log
(N$^2$LL) approximation 
(also $g_3(\lambda)$), and so on.
For the new, long-distance form factor $S_L$,
we have defined an analog 
logarithmic-approximation scheme: 
Leading-Log massive
(LL$_m$) accuracy, involving
only the leading function $f_1(\rho)$
($h_1(\rho)\equiv 0$);
Next-to-Leading-Log massive
(NLL$_m$) accuracy, including
also the functions $f_2(\rho)$ and $h_2(\rho)$;
Next-to-Next-to-Leading-Log massive
(N$^2$LL$_m$) accuracy, involving
also $f_3(\rho)$ and $h_3(\rho)$, and so on.

In eqs.(\ref{eq_define_gs}) and
(\ref{eq_define_g2}) we have explicitly
computed the functions $f_1(\rho)$, $f_2(\rho)$
and $f_3(\rho)$,
together with the functions $h_2(\rho)$
and $h_3(\rho)$, entering the mass-dependent, long-distance form
factor at N$^2$LL$_m$ accuracy.

At Next-to-Leading Logarithmic massive (NLL$_m$) accuracy, the subleading soft effects,
related to the coefficient $D_1$,
can be interpreted as a rescaling
of the hard scale $Q$ 
of the process,
\beq
\label{eq_rescale}
\textrm{NLL}_m:
\,\,\, Q \quad \Rightarrow \quad
\tilde{Q} \, \equiv \, Q \, \exp\left( \frac{D_1}{2 \, A_1} \right)
\, = \, \frac{Q}{\sqrt{e}}. 
\eeq
The inverse of the rescaling factor on the 
last member of eq.(\ref{eq_rescale}) can be interpreted
as the NLL$_m$ correction to the 
(classical) dead-cone opening angle:
\beq
\textrm{NLL}_m:
\,\,\,
\theta \, \gsim \, \frac{1}{\gamma}
\quad \Rightarrow \quad
\theta \, \gsim \, 
\frac{\sqrt{e}}{\gamma};
\eeq
where $\sqrt{e} \simeq 1.64872$
and $\gamma \equiv 1/\sqrt{1-v^2} = E/m$ is the Lorenz factor of the heavy quark,
with $v\equiv ds/dt$ its ordinary 3-velocity ($c=1$).

However, beyond NLL$_m$ approximation, the above interpretation
of the subleading soft $D$-terms as a rescaling
effect of the hard scale or the
dead-cone opening angle, {\it breaks down}.
At N$^2$LL$_m$, for example, one has
to introduce {\it two} different rescaling factors
(rather than one),
entering different contributions to the form factor.
In even higher orders, a progressively
larger number of different rescaling factors
has to be introduced.


\end{document}